\documentstyle[12pt]{article}

\catcode`\@=11
\def\marginnote#1{}

\newcount\hour
\newcount\minute
\newtoks\amorpm
\hour=\time\divide\hour by60
\minute=\time{\multiply\hour by60 \global\advance\minute
by-\hour}\edef\standardtime{{\ifnum\hour<12
\global\amorpm={am}%
        \else\global\amorpm={pm}\advance\hour by-12 \fi
        \ifnum\hour=0 \hour=12 \fi
        \number\hour:\ifnum\minute<10
0\fi\number\minute\the\amorpm}}
\edef\militarytime{\number\hour:\ifnum\minute<10
0\fi\number\minute}

\def\draftlabel#1{{\@bsphack\if@filesw {\let\thepage\relax
   \xdef\@gtempa{\write\@auxout{\string
      \newlabel{#1}{{\@currentlabel}{\thepage}}}}}\@gtempa
   \if@nobreak \ifvmode\nobreak\fi\fi\fi\@esphack}
        \gdef\@eqnlabel{#1}}
\def\@eqnlabel{}
\def\@vacuum{}
\def\draftmarginnote#1{\marginpar{\raggedright\scriptsize\tt#1}}
\def\draft{\oddsidemargin -.5truein
        \def\@oddfoot{\sl preliminary draft \hfil
        \rm\thepage\hfil\sl\today\quad\militarytime}
        \let\@evenfoot\@oddfoot \overfullrule 3pt
        \let\label=\draftlabel
        \let\marginnote=\draftmarginnote

\def\@eqnnum{(\theequation)\rlap{\kern\marginparsep\tt\@eqnlabel}%
\global\let\@eqnlabel\@vacuum}  }

\def\numberbysection{\@addtoreset{equation}{section}
        \def\theequation{\thesection.\arabic{equation}}}

\def\underline#1{\relax\ifmmode\@@underline#1\else
 $\@@underline{\hbox{#1}}$\relax\fi}

\catcode`@=12
\relax

\numberbysection

\advance \topmargin by -\headheight
\advance \topmargin by -\headsep

\evensidemargin \oddsidemargin
\def\sect#1{\section{#1}}

\def\rf#1{(\ref{#1})}
\def\lab#1{\label{#1}}
\def\nonu{\nonumber}
\def\br{\begin{eqnarray}}
\def\er{\end{eqnarray}}
\def\be{\begin{equation}}
\def\ee{\end{equation}}

\def\({\left(}
\def\){\right)}

\newcommand{\ct}[1]{\cite{#1}}
\newcommand{\bi}[1]{\bibitem{#1}}

\relax

\def\a{\alpha}

\def\b{\beta}

\def\d{\delta}

\def\g{\gamma}

\def\h{{1\over 2}}

\def\o{\over}

\def\pa{\partial}
\def\pr{\prime}

\def\ra{\rightarrow}
\def\s{\sigma}

\def\tp0{\Theta_{+}^{(0)}}
\def\tm0{\Theta_{-}^{(0)}}

\def\vp{\varphi}

\def\f#1#2#3 {f^{#1#2}_{#3}}

\def\win1{{\sf w_{1+\infty}}}

\def\Win1{{\sf W_{1+\infty}}}

\def\rlx{\relax\leavevmode}
\def\inbar{\vrule height1.5ex width.4pt depth0pt}
\def\IZ{\rlx\hbox{\sf Z\kern-.4em Z}}
\def\IR{\rlx\hbox{\rm I\kern-.18em R}}
\def\IC{\rlx\hbox{\,$\inbar\kern-.3em{\rm C}$}}
\def\IN{\rlx\hbox{\rm I\kern-.18em N}}
\def\IO{\rlx\hbox{\,$\inbar\kern-.3em{\rm O}$}}
\def\IP{\rlx\hbox{\rm I\kern-.18em P}}
\def\IQ{\rlx\hbox{\,$\inbar\kern-.3em{\rm Q}$}}
\def\IF{\rlx\hbox{\rm I\kern-.18em F}}
\def\IG{\rlx\hbox{\,$\inbar\kern-.3em{\rm G}$}}
\def\IH{\rlx\hbox{\rm I\kern-.18em H}}
\def\II{\rlx\hbox{\rm I\kern-.18em I}}
\def\IK{\rlx\hbox{\rm I\kern-.18em K}}
\def\IL{\rlx\hbox{\rm I\kern-.18em L}}
\def\one{\hbox{{1}\kern-.25em\hbox{l}}}
\def\0#1{\relax\ifmmode\mathaccent"7017{#1}%
B        \else\accent23#1\relax\fi}

\def\PRL#1#2#3{{\sl Phys. Rev. Lett.} {\bf#1} (#2) #3}
\def\ibid#1#2#3{{\sl ibid.} {\bf#1} (#2) #3}
\def\NPB#1#2#3{{\sl Nucl. Phys.} {\bf B#1} (#2) #3}

\def\CMP#1#2#3{{\sl Commun. Math. Phys.} {\bf #1} (#2) #3}
\def\PRD#1#2#3{{\sl Phys. Rev.} {\bf D#1} (#2) #3}

\def\PLB#1#2#3{{\sl Phys. Lett.} {\bf #1B} (#2) #3}

\def\PTP#1#2#3{{\sl Prog. Theor. Phys.} {\bf #1} (#2) #3}
\def\SPTP#1#2#3{{\sl Suppl. Prog. Theor. Phys.} {\bf #1} (#2) #3}
\def\AoP#1#2#3{{\sl Ann. of Phys.} {\bf #1} (#2) #3}

\def\PR#1#2#3{{\sl Phys. Reports} {\bf #1} (#2) #3}

\def\IJMPA#1#2#3{{\sl Int. J. Mod. Phys.} {\bf A#1} (#2) #3}
\def\IJMPB#1#2#3{{\sl Int. J. Mod. Phys.} {\bf B#1} (#2) #3}

\def\TMP#1#2#3{{\sl Theor. Mat. Phys.} {\bf #1} (#2) #3}

\def\MPLA#1#2#3{{\sl Mod. Phys. Lett.} {\bf A#1} (#2) #3}

\begin{document}
\begin{titlepage}
\vspace*{-1cm}
\hyphenation{co-lors}
\hyphenation{non-va-ni-shing}
\vspace{.2in}
\begin{center}
{\large\bf Reduction of affine two-loop WZNW model, the Toda system coupled to the matter and topological confinement}
\end{center}

\vspace{1 in}

\begin{center}
Harold Blas

\vspace{.5 cm}
\small

\par \vskip .1in \noindent
Instituto de F\'\i sica Te\'orica - IFT/UNESP\\
Rua Pamplona 145\\
01405-900  S\~ao Paulo-SP, BRAZIL\\

\normalsize
\end{center}

\vspace{1.5in}

\begin{abstract}

The conformal affine Toda model coupled to the matter field (CATM) is obtained through a classical reduction of the $sl(2)^{(1)}$ affine two-loop WZNW model. After spontaneously broken the conformal symmetry by means of BRST analysis, we end up with an effective theory, the so called affine Toda model coupled to the matter (ATM). Further, using a bosonization technique we recover from this theory the sine-Gordon model plus a free massless scalar field. The ATM model is considered as a QCD-motivated integrable field theory, since it describes various features in the baryonic sector of the low-energy effective Lagrangian of QCD in two dimensions with one flavor and two  colors. Imposing the equivalence of the Noether and topological currrents as a constraint, it is shown that the intercharge ``quark''- ``anti-quark'' static potential reveals a linear confinement behavior for large intercharge separation.

\end{abstract}

\vspace{1 cm} 


\end{titlepage}

\sect{Introduction}
\label{sec:intro}

Integrable theories in two-dimensions have been an extraordinary laboratory for the understanding of basic nonperturbative aspects of physical theories and  various phenomena  such as  dynamical mass generation,
asymptotic freedom, and quark confinement, relevant in more realistic
models, have been tested. In QCD$_{2}$ one of the open questions is related to the higher order correction to the limit $m/e\rightarrow 0$ ($m$, quark mass and $e$, coupling constant). A speculation regarding this issue was set forward earlier (see, for example, Ref. \ct{sonnenschein}), that the low-energy effective action of QCD$_{2}$ might be related to some massive two dimensional integrable models, thus leading to the exact solution (not semi-classical) of the strong coupled  QCD$_{2}$. Although some hints toward a possible integrable structure in QCD$_{2}$ have been encountered the problem still remains largely open (see, e.g., \ct{alves}). The problem resides, for example, in a better understanding of the meson spectrum of the theory, for finite number of colors ($N_{c}$). It is believed that integrability of a theory implies stability of its bound states (mesonic spectrum). Therefore, one could expect vanishing decay amplitudes for the mesonic states of the theory. Several numerical computations indicate the contrary \ct{moha}.     

In the present paper we study the recently proposed conformal affine $sl(2)^{(1)}$ Toda model coupled to (Dirac) matter field (CATM) which is an example of a wide class of integrable theories presented in \ct{matter}. The zero curvature representation, the construction of the general solution and many other properties are discussed in \ct{matter}. This model possesses a Noether current depending only on the matter field and under some circunstances, it is possible to choose one solution in each orbit of the conformal group such that, for these solutions, the $U(1)$ current is equal to a topological current depending only on the Toda field. Such equivalence leads, at the classical level, to the localization of the (Dirac) matter field inside the Toda field soliton. It is well known the relevance of localized
classical solutions of non-linear relativistic field equations to the
corresponding quantum theories. In particular, solitons can be associated with quantum
extended-particle states; this picture is valid for the soliton of our model, which at the classical level is associated to the Toda field $\vp$. Then one may regard the model a sort of one dimensional bag model for QCD. Besides, an additional feature is present; the masses of solitons and particles are proportional to the $U(1)$ Noether charge. This fact indicates the existence of a sort of duality in these models involving solitons and particles \ct{montonen}. The interest in such models comes from their integrability and duality properties \ct{matter, bla1}, which can be used as a toy model to understand the electric-magnetic duality in four dimensional gauge theories, conjectured in \ct{montonen} and developed in \ct{vw}. Thereby nonperturbative analysis of the spectrum and of the phase structure in SUSY Yang-Mills theory becomes possible. 

The CATM theory is well defined mathematically for a set of fields which, in general, may be complex fields giving rise to a complex Lagrangian. By imposing convenient reality conditions on the fields of the model it is possible to define an integrable sub-model with a real Lagrangian. Moreover, there exists a related off-critical model, the so called affine $sl(2)^{(1)}$ Toda model coupled to  the matter (ATM), which can be obtained at the classical or quantum mechanical level through some convenient reduction process starting from the CATM model \ct{bla1, bla}. Recently the classical one and two soliton solutions of the model have been found, as well as the time-delays arising from the collisions of two solitons, and the implications of the reality conditions (imposed on the fields of the CATM model) on the solitonic solutions have been studied \ct{bla1}. In addition, in Ref. \ct{bla1} it was suggested the possibility of implementing a confinement mechanism of some degrees of freedom of the ATM model inside the baryons (solitons) of the sine-Gordon type reduced model. The symplectic structure of the off-critical ATM model has recently been studied \ct{bla}. It was performed in the context of Faddeev-Jackiw and (constrained) symplectic methods (Barcelos-Neto, Montani and Wotzasek); by imposing the equivalence between the Noether and topological currents as a constraint, the authors have been able to obtain either, the sine-Gordon model or the massive Thirring model, through a Hamiltonian reduction and gauge fixing the symmetries of the model in two different ways.

Here we perform the reduction process of the CATM model to the off-critical ATM model through a BRST analysis and study the quantum spectrum of this reduced model. By bosonizing the spinor field we show that the off-critical ATM model is equivalent to a theory of a free massless scalar and a sine-Gordon field. A further quantum reduction process is implemented, this time to eliminate from the spectrum the modes associated to the free massless scalar; this amounts to imposing the equivalence between the Noether and topological currents as a constraint. Then it is shown that in the solitons (baryons) of the sine-Gordon theory the confinement of some degrees of freedom of the ATM model does take place. 

Moreover, by means of a semi-classical analysis we find the ``color'' intercharge potential in the ATM model, obtaining a linearly rising potential for large intercharge separation, thus revealing the presence of a confining phase in the model. This behaviour leads to an interesting analogy with what one expects to happen in QCD. Then we suggest an interesting connection of the ATM model with the baryonic sector of the low-energy effective action of QCD$_{2}$ with one flavor ($N_{f}=1$) and two colors ($N_{c}=2$) \ct{ferrando2}. With that motivation we reproduce the bosonized form of QCD$_{2}$ in Baluni's gauge \ct{baluni}. This gauge gives a convenient framework to study the topological realization of color symmetry in QCD$_{2}$, and then reproduces the bosonized form of the theory in a form adequate to our discussion in connection to the bosonized formulation of the ATM model. 
      
The paper is organized as follows. In section \ref{sec:catm} we present some relevant aspects of the conformal affine $sl(2)^{(1)}$ Toda model coupled to matter (Dirac) field. Section \ref{sec:wznw} presents a Hamiltonian reduction of the two-loop WZNW model to obtain the equations of motion of the CATM theory. Section \ref{sec:real} presents a submodel in which a suitable reality conditions on the fields of CATM are imposed so as to define positive-definite kinetic term in the scalar sector and a usual spinorial kinetic term of the Lagrangian. These conditions will give rise to an off-critical model (ATM) with real Lagrangian, and the relevant Hamiltonian bounded from below. Also for the sake of completeness, we reproduce some relevant steps of the classical reduction process: CATM $\rightarrow$ ATM \ct{bla}. Moreover we present the general soliton (antisoliton) solution of the real Lagrangian submodel. In subsection \ref{subsec:masses} we present the masses of the fundamental particles and solitons, using a linear field approximation of the equations of motion. 
 Section \ref{sec:quantum} deals with the quantum aspects of the model. In subsection \ref{subsec:brst} through a BRST analysis the conformal symmetry of the CATM is spontaneously broken, then defining the off-critical ATM model. In subsection \ref{subsec:boso}, using bosonization techniques we arrive at the sine-Gordon model (massive Thirring) plus a free massless scalar field. A further quantum reduction is performed imposing the currents (Noether and topological) equivalence, then obtaining the sine-Gordon (massive Thirring) model only. In section \ref{sec:color} we present the realization of color symmetry in QCD$_{2}$ bosonizing the theory such that the topological confinement mechanism is more transparent.
In section \ref{sec:topo} a confinement mechanism of the `color' sector of the ATM model inside the soliton (baryon) is explained observing the behaviour of the static intercharge potential.
 
Finally, section \ref{sec:discussions} summarise the main point and contains some remarks about the results. The appendix \ref{appa} provides the relevant notations and conventions, and appendix \ref{appb} presents some useful results concerning the $sl(2)^{(1)}$ affine Lie algebra.

\section{The conformal affine Toda model coupled to the matter (CATM): the principal gradation of $sl(2)^{(1)}$ case}
\label{sec:catm}

We will be interested in the so called conformal affine Toda system coupled to the matter field, whose
construction, in the particular case of the algebra $sl(2)^{(1)}$, using the
parlance of the original reference, will be summarized in the following
paragraphs.

We discuss the example associated with the principal gradation of the
untwisted affine Kac-Moody algebra $sl(2)^{(1)}$ \footnote{In Ref. \ct{halpern} the terminology ``affine Lie algebra'' is preferred. I would like to thank Professor M.B. Halpern for correspondence, and for pointing out to me this terminology used in physics.}. This belongs to a special
class of models introduced in \ct{matter} possessing a $U(1)$ Noether current
depending only on the matter fields. It is then possible, under some
circunstances, to choose one solution in each orbit of the conformal group,
such that for these solutions, that $U(1)$ current is proportional to a
topological current depending only on the (gauge) zero grade field.

The zero curvature condition in light-cone coordinates $x_{\pm }=t\pm x$
takes the form
\br
\lab{zerocur}
\partial_{+}A_{-}-\partial_{-}A_{+}+\left[ A_{+}\,,\,
A_{-}\right] =0.
\er

The connections are of the form 
\br
\lab{conn}
A_{+}=-BF^{+}B^{-1},\qquad A_{-}=-\partial _{-}BB^{-1}+F^{-},
\lab{eq4}
\er
where the mapping $B$ is parametrised as 
\begin{equation}
B=be^{\nu C}e^{\eta Q_{{\bf s}}}=e^{\varphi H^{0}}e^{\widetilde{\nu }C}e^{\eta
Q_{{\bf s}}},\,\,\, \mbox{with}\,\,\, b=e^{\varphi \widetilde{H}^{0}},
\lab{Bmapp}
\end{equation}
and so $\widetilde{\nu }=\nu -\frac{1}{2}\varphi .$

The special class of models with the $U(1)$  Noether current
proportional to a topological current, occurs for those models where the grades of the first constant terms of the nonvanishing grade potentials $F^{\pm }$ are equal to $\pm N_{{\bf s}}$, respectively. So, the potentials $F^{\pm }$ are of the form
\br
\lab{FF1}
F^{+}\equiv E_{2}+F_{1}^{+},\qquad F^{-}\equiv E_{-2}+F_{1}^{-}, 
\er
where 
\br
\lab{FF2}
F_{1}^{+}=2\sqrt{im}(\psi _{R}E_{+}^{0}+\widetilde{\psi }_{R}E_{-}^{1}),\,\,\,
F_{1}^{-}=2\sqrt{im}(\psi _{L}E_{+}^{-1}-\widetilde{
\psi }_{L}E_{-}^{0}), 
\er
and $E_{\pm 2}=mH^{\pm 1}$ ($m$ =constant).

Substituting the gauge potentials \rf{conn} and \rf{FF1} into \rf{zerocur}, one gets the equations of motion 
\br
\lab{poteqn1}
\partial _{+}(\partial _{-}bb^{-1})+\partial _{+}\partial _{-}\nu C
&=&m^{2}e^{2\eta }[H^{-1},bH^{1}b^{-1}]+e^{\eta
}[F_{1}^{-},bF_{1}^{+}b^{-1}], \\
\lab{poteqn2}
\partial _{-}F_{1}^{+} &=&me^{\eta }[H^{1},b^{-1}F_{1}^{-}b], \\
\lab{poteqn3}
\partial _{+}F_{1}^{-} &=&-me^{\eta }[H^{-1},bF_{1}^{+}b^{-1}]. 
\er

Making use of the explicit form of the connections $b$ and $F_{1}^{\pm}$, Eqs.\rf{Bmapp} and \rf{FF2}, respectively, into Eqs.\rf{poteqn1}-\rf{poteqn3}, one can get the equations of motion
\begin{eqnarray}
\lab{eqn1}
\partial ^{2}\varphi &=&-4m_{\psi }\overline{\psi }\gamma _{5}e^{\eta
+2\varphi \gamma _{5}}\psi ,  \\
\lab{eqn2}
\partial ^{2}\widetilde{\nu } &=&-2m_{\psi }\overline{\psi }(1-\gamma
_{5})e^{\eta +2\varphi \gamma _{5}}\psi -\frac{1}{2}m_{\psi }^{2}e^{2\eta },
 \\
\lab{eqn3}
\partial ^{2}\eta &=&0,  \\
\lab{eqn4}
i\gamma ^{\mu }\partial _{\mu }\psi &=&m_{\psi }e^{\eta +2\varphi \gamma
_{5}}\psi ,  \\
\lab{eqn5}
i\gamma ^{\mu }\partial _{\mu }\widetilde{\psi } &=&m_{\psi }e^{\eta
-2\varphi \gamma _{5}}\widetilde{\psi },\,\,\,m_{\psi }\equiv 4m.
\end{eqnarray}

The equations \rf{eqn1}-\rf{eqn5} can be derived from the Lagrangian
\br
\frac{1}{k}{\cal L}_{CATM}\, =\, {1\o 4} \pa_{\mu} \vp \, \pa^{\mu} \vp
+ \h  \pa_{\mu} \nu \, \pa^{\mu} \eta
- {1\o 8}\, m_{\psi}^2 \, e^{2\,\eta} 
+ i  {\bar{\psi}} \gamma^{\mu} \pa_{\mu} \psi
- m_{\psi}\,  {\bar{\psi}} \,
e^{\eta+2\vp\,\gamma_5}\, \psi.
\lab{lagrangian0}
\er

It is real (for $\eta $ $=$real) if $\widetilde{\psi }$ is proportional to the
complex conjugate of $\psi$, and if $\varphi $ is pure imaginary. This is true for the particular solutions of \rf{eqn1}-\rf{eqn5} such as the 1-soliton (1-antisoliton),
soliton-soliton (antisoliton-antisoliton) \ct{bla1}. The general reality conditions, imposed on the fields in order to define a real Lagrangian with its corresponding Hamiltonian bounded from below, will be discussed in section \ref{sec:real}.

\section{Affine two-loop WZNW models and reduction}
\label{sec:wznw}

We provide a brief presentation of the Hamiltonian reduction to obtain the  conformal affine Toda model coupled to the matter field (CATM) starting from the two-loop WZNW theory. The action for the two-loop WZNW model is \ct{twoloop, martellini} 
\br
S_{WZNW}(\hat{g})=-\frac{\kappa}{8\pi}\int_{\pa {\bf B}}Tr(\hat{g}^{-1}\pa_{\mu}\hat{g}\hat{g}^{-1}\pa^{\mu}\hat{g})+\frac{\kappa}{12\pi}\int_{{\bf B}}\epsilon^{\mu\nu\sigma}Tr(\hat{g}^{-1}\pa_{\mu}\hat{g}\hat{g}^{-1}\pa_{\nu}\hat{g}\hat{g}^{-1}\pa_{\sigma}\hat{g})
\er
with the fields being mappings from ${\bf B}$ (${\bf B}=D \mbox{x} {\bf R}$, with $D$ being a disc) to the affine KM group $\hat{G}$. 

The equations of motion read
\br
\lab{LR}
\pa_{+}(\pa_{-}\hat{g}\hat{g}^{-1})=0, \,\,\,\,\,\pa_{-}(\hat{g}^{-1}\pa_{+}\hat{g})=0.
\er

We now consider those group elements that can be written in the modified Gauss decomposition form
\br
\hat{g}=N B M,
\er
where N, B and M are spanned by the subalgebras $\hat{\cal{G}}_{+}$, $\hat{\cal{G}}_{0}$ and $\hat{\cal{G}}_{-}$, respectively. Introducing the mappings $K_{L/R}$ as $\pa_{-}\hat{g}\hat{g}^{-1}=NK_{L}N^{-1}$ and  $\hat{g}^{-1}\pa_{+}\hat{g}=M^{-1}K_{R}M$, and taking into account \rf{LR} one gets
\br
\lab{KL}
\pa_{-}K_{R}=-\left[K_{R},\, \pa_{-}MM^{-1}\right],\,\,\, \pa_{+}K_{L}=-\left[K_{L},\, N^{-1}\pa_{-}N\right].
\er  

To recover the model \rf{poteqn1}-\rf{poteqn3} we impose the constraints
\begin{eqnarray}
\lab{constr1}
&(\pa_{-}MM^{-1})_{-2}=B^{-1}E_{-2}B,\,\,(\pa_{-}MM^{-1})_{<-2}=0,\\
\lab{constr2}
&(N^{-1}\pa_{+}N)_{2}=BE_{2}B^{-1}, \,\,(N^{-1}\pa_{+}N)_{>2}=0.
\end{eqnarray}

Then, the mappings $\pa_{-}MM^{-1}$ and $N^{-1}\pa_{+}N$ have components in the subspaces $\hat{\cal G}_{-1}$ and $\hat{\cal G}_{1}$, respectively. To give a relation to system \rf{poteqn1}-\rf{poteqn3}, make the correspondence
\br
B\pa_{-}MM^{-1}B^{-1}\,=\,E_{-2}+F_{1}^{-}\equiv F^{-},\,\,\,B^{-1}N^{-1}\pa_{+}NB\,=\,E_{2}+F_{1}^{+}\equiv F^{+}.
\er

Using the definitions given for $K_{L/R}$ and taking into account the constraints \rf{constr1}-\rf{constr2}, we can substitute them into \rf{KL}, showing that $B, F^{+}$ and  $F^{-}$ satisfy \rf{poteqn1}-\rf{poteqn3}. Then, the conformal and integrable model called affine Toda model coupled to the matter field (CATM) can  be obtained from the affine (two-loop) WZNW model by means of a Hamiltonian reduction. This relation of the CATM theory to the one of WZNW model allows us to write the following relationship between their coupling constants
\br
\lab{couplings}
\kappa\, =\, 2\pi k.
\er

Let us recall that in the WZNW model it is a well known fact that the coupling constant $\kappa$ takes integer values.

\section{A real Lagrangian sub-model}
\label{sec:real}

From the point of view of their eventual quantization it would be important to distinguish those models whose kinetic terms are positive-definite and whose action is real. So, we are going to consider a sub-model of the conformal affine Toda system coupled to matter (Dirac) field (CATM) \rf{lagrangian0}, defining the two-dimensional field theory \ct{bla1, bla} 
\br
 \frac{1}{k}{\cal L} = {1\o 4} \pa_{\mu} \vp \, \pa^{\mu} \vp
+ \h  \pa_{\mu} \nu \, \pa^{\mu} \eta
+ {1\o 8}\, m_{\psi}^2 \, e^{2\,\eta} 
+ i  {\bar{\psi}} \gamma^{\mu} \pa_{\mu} \psi
- m_{\psi}\,  {\bar{\psi}} \,
e^{\eta+2i\vp\,\gamma_5}\, \psi,
\lab{lagrangian}
\er
where ${\bar{\psi}} \equiv {\psi}^{\dagger} \,\gamma_0$, and $\varphi$, $\eta $ and $\nu$ are real fields. 

The Lagrangian \rf{lagrangian} differs from the Lagrangian ${\cal L}_{CATM}$ presented in \rf{lagrangian0} in three points:
 
i) the conformal affine Toda system coupled to matter (Dirac) field (CATM), Eq. \rf{lagrangian0}, contains two Dirac spinor fields, $\widetilde{\psi}$ and $\psi$, and a complex $\vp$ field, as well as, the real fields $\nu$ and $\eta$.

ii) we have imposed the reality conditions, $\widetilde{\psi}=-\psi^{*}$ (the star means complex conjugation) and $\vp$ pure imaginary (making the replacement $\vp \rightarrow i\vp$ in \rf{lagrangian0}, we have a real $\vp$ in \rf{lagrangian}), in order to have a real Lagrangian. Moreover, for later convenience, we have made the change $\nu \rightarrow -\nu$. The implications of these reality conditions on the solitonic solutions of the model \rf{lagrangian0} were studied in \ct{bla1}.

iii) an overall minus sign comes out in order to construct a Hamiltonian bounded from below.

The point ii) deserves a far greater attention. The classical theory defined by  \rf{lagrangian0} is well defined mathematically, even for a general complex nature of the field solutions (complex Lagrangian). This fact immediately prompts the reaction that the Hamiltonian, hence the energy of configurations of such a system, can not be bounded below, spelling disaster both at the classical level (unstable modes) and at the quantum level (loss of unitarity). This issue certainly has to be handled carefully before this type of model can be considered as providing a sound toy-model for aspects of dualities between particles and solitons, as is one of the motivations for studying it. Here we follow the prescription to restrict the model to a subspace of well-behaved classical solutions. For example the one and two soliton (anti-soliton) solutions satisfy the above reality conditions \ct{bla1}. Then taking into account the overall minus sign and the above reality conditions allow us to define a sensible physical Lagrangian. These kind of issues in the case of affine Toda field theories are discussed in Refs. \ct{hermitian}, and for non-abelian Toda theories see, for example, Ref. \ct{miramontes}.   

Moreover, instead of point ii), the following equally meaninful reality conditions could have been imposed

ii$^{\prime}$) $\widetilde{\psi}=\psi^{*}$, $\vp \rightarrow i\vp - i\pi/2$ and $\nu \rightarrow -\nu$, supplied with the change $x^{\mu} \rightarrow -x^{\mu}$. These reality conditions could have also allowed us to end up with Lagrangian \rf{lagrangian} starting from the CATM Lagrangian \rf{lagrangian0}, provided point iii) were also taken into account. 

The most general, $\widetilde{\psi}=e_{\psi}\psi^{*}$ and $\vp$ pure imaginary, reality conditions and their implications on the solitonic solutions of this model were also presented in \ct{bla1}. See below (subsection \ref{subsec:masses}) a discussion about the implications of these reality conditions on the classical masses of the fundamental particles and the particle/soliton dualities of the theory. 

We are going to mention the relevant symmetries of \rf{lagrangian}. The model \rf{lagrangian} is invariant under the conformal transformations
\be
\lab{conformal}
x_{+} \ra {\hat x}_{+} = f(x_{+}) \, , \qquad 
x_{-} \ra {\hat x}_{-} = g(x_{-}),
\lab{ct}
\ee
with $f$ and $g$ being analytic functions; and with the fields transforming 
as
\br
\nonu
\vp (x_{+}\, , \, x_{-}) &\ra& 
{\hat {\vp}}({\hat x}_{+}\, , \,  {\hat x}_{-}) = 
\vp (x_{+}\, , \, x_{-}) \, ,
\\
e^{-\nu (x_{+}\, , \, x_{-})} &\ra& 
e^{-{\hat \nu}({\hat x}_{+}\, , \, 
{\hat x}_{-})} = \( f^{\pr}\)^{\d} \, \( g^{\pr}\)^{\d}
e^{-\nu (x_{+}\, , \, x_{-})} \, ,
\lab{ctf}\\
\nonu
e^{-\eta (x_{+}\, , \, x_{-})} &\ra& e^{-{\hat \eta}({\hat x}_{+}\, , \, 
{\hat x}_{-})} = \( f^{\pr}\)^{\h} \, \( g^{\pr}\)^{\h}  e^{-\eta (x_{+}\, 
, \, x_{-})} \, ,
\\ 
\nonu
\psi (x_{+}\, , \, x_{-}) &\ra & {\hat {\psi}} ({\hat x}_{+}\, , \, 
{\hat x}_{-}) =   e^{{1\o 2}\( 1+ \gamma_5\) \log \( f^{\pr}\)^{-\h} 
+ {1\o 2}\( 1- \gamma_5\) \log \( g^{\pr}\)^{-\h}}
\, \psi (x_{+}\, , \, x_{-}) \, ,
\er
where the conformal weight $\d$, associated to $e^{-\nu}$, is arbitrary.

On the other hand, \rf{lagrangian} is also invariant under the commuting $U(1)_L \otimes U(1)_R$ left and right local gauge transformations
\br
\lab{leri1}
\vp \ra \vp + \xi_{+}\( x_{+}\) + \xi_{-}\( x_{-}\) \; ; \qquad 
\nu \ra \nu \; ; \qquad \eta \ra \eta 
\er
and
\br
\lab{leri2} 
\psi \ra e^{- i\( 1+ \gamma_5\) \xi_{+}\( x_{+}\) 
+ i\( 1- \gamma_5\) \xi_{-}\( x_{-}\)}\, \psi.  
\er

By a special choice of $\xi_{+}\( x_{+}\) = - \xi_{-}\( x_{-}\) = - \h
\theta$, with $\theta = {\rm const.}$, one gets a global $U(1)$ transformation 
\be
\vp \ra \vp  \; , \qquad 
\nu \ra \nu \; , \qquad \eta \ra \eta   \; , \qquad 
\psi \ra e^{i \theta} \, \psi  \;, 
\lab{globalu1}
\ee
and the Noether current, associated to the choice, is given by
\be
J^{\mu} = {\bar{\psi}}\, \gamma^{\mu}\, \psi \, , \qquad
\pa_{\mu}\, J^{\mu} = 0.
\lab{noethersl2}
\ee

Another choice could also be possible by taking $\xi_{+}\( x_{+}\) = \xi_{-}\( x_{-}\) = - \h\alpha$, with $\alpha = {\rm const.}$. In this way, one has the global chiral symmetry
\be
\lab{chiralglobal}
\psi \ra e^{i\gamma_5 \a}\, \psi \; , \qquad 
\vp \ra \vp -  \a \; , \qquad 
\nu \ra \nu \; , \qquad \eta \ra \eta,
\ee
and the corresponding Noether current is 
\be
J_5^{\mu} =  \bar \psi \gamma^\mu \gamma_5 \psi 
+{1\over 2} \partial^\mu \vp  
\; , \qquad \qquad \pa_{\mu}J_5^{\mu} =0. 
\lab{chiral}
\ee

Concerning the topological current, the Lagrangian \rf{lagrangian} is invariant under $\vp \ra \vp + n \pi$, with all the other fields unchanged. Indeed, the vacua are infinitely degenerate, and the topological charge 
\be
Q_{\rm topol.} \equiv \int \, dx \, j^0 
\, , \qquad
j^{\mu} =  {1\o{2\pi}}\epsilon^{\mu\nu} \pa_{\nu} \, \vp, 
\lab{topological}
\ee
depending only on the asymptotic values of $\vp$, at $x=\pm \infty$, can take non-zero values.
 
Next, for the sake of completeness, we are going to discuss the relevance of these symmetries in the reduction process to the off-critical affine Toda model coupled to the matter (ATM) \ct{bla}. Associated to the conformal symmetry \rf{conformal} there are two chiral currents 
\be
{\cal J}= - i{\psi}^{\dagger} \( 1+\gamma_5 \)\psi 
+i \partial_+\vp +\partial_+\eta, \qquad
{\bar {\cal J}}=  i{\psi}^{\dagger} \( 1-\gamma_5 \)\psi 
+i \partial_-\vp+ \partial_-\eta,
\lab{chiralcur}
\ee
satisfying 
\be
\partial_-{\cal J}=0 \; ; \qquad \quad  \partial_+{\bar {\cal J}}=0.
\ee

Notice, from \rf{ctf}, that the currents ${\cal J}$ and ${\bar {\cal J}}$ have
conformal weights $(1,0)$ and $(0,1)$ respectively. Under the conformal transformations \rf{conformal}, the chiral currents transform as
\br
\lab{chi1}
\;\;{\cal J}(x_{+})&\longrightarrow& [\ln f^{\pr}( x_{+})
]^{-1}\left({\cal J}(x_{+})-[\ln f^{\pr}(x_{+})]^{\pr}\right),\\
\lab{chi2}
\overline{{\cal J}}(x_{-})&\longrightarrow& [\ln g^{\pr}(x_{-})
]^{-1}\left(\overline{{\cal J}}(x_{-})-[\ln
g^{\pr}( x_{-})]^{\pr}\right).
\er 
 
Then, given a solution of the model, one can always map it, under a conformal transformation, into a solution where

\be
{\cal J}=0 \; , \qquad \quad {\bar {\cal J}}=0. 
\lab{constraints}
\ee

Such a procedure amounts to ``gauging'' away the free field $\eta$. In other words, \rf{constraints} are constraints implementing a Hamiltonian reduction. This  resembles the connection between the affine (AT) and conformal affine Toda (CAT) models performed in \ct{aratyn, const} through a gauge fixing of the conformal symmetry. The quantum version of the reduction CAT $\rightarrow$ AT was performed in  \ct{bonora}.

Now, let us explain the gauge fixing procedure. Notice from \rf{ctf} that $\vp$ is a scalar under conformal transformations and if we set $\delta$ to zero, $e^{\nu}$ is also scalar. On the other hand $e^{\eta}$ is a $(1/2,1/2)$ primary field. Let us perform a conformal transformation \rf{ctf} with
\br
f^{\pr}(x_{+})=e^{2\eta_+(x_+)},\,\,\,\,g^{\pr}(x_{-})=e^{2\eta_-(x_-)},
\er    
where $\eta_{\pm}(x_{\pm})$ are solutions of the $\eta$ free field ( $\eta(x_+,x_-)=  \eta_+(x_+)+\eta_-(x_-)$;\, with  $\eta_{\pm}(x_\pm)$, being arbitrary functions), one gets
\br
\hat{\vp} (\hat{x}_{+}\, , \, \hat{x}_{-}) \ra {{\vp}}(x_{+},x_-),\,\,\, e^{-\hat{\eta}_+(\hat{x}_+,\hat{x}_-)}\ra 1,\,\,\, e^{-\hat{\nu}_+(\hat{x}_+,\hat{x}_-)}\ra e^{-\nu_+(x_+,x_-)}. 
\er

Therefore, we are choosing one solution in each orbit of the conformal group. Then for every regular solution of the $\eta$ field the CATM defined on a space-time $(x_+,x_-)$ corresponds to an off-critical sub-model which we call the affine Toda model coupled to the matter (ATM), with the extra field $\nu$ defined on a space time $(\hat{x}_+,\hat{x}_-)$. For  the particular solution $\eta = 0$ the CATM and the ATM are defined on the same space-time.
 
Thus setting the $\eta$ field to zero in the equations of motion of the CATM model \rf{eqn1}-\rf{eqn5} we recover the equations (for the spinor $\psi$ and scalar $\vp$ fields),  
\begin{eqnarray}
\lab{eqna1}
\partial ^{2}\varphi &=&-4m_{\psi }\overline{\psi }\gamma _{5}e^{2\varphi \gamma _{5}}\psi ,  \\
\lab{eqna2}
i\gamma ^{\mu }\partial _{\mu }\psi &=&m_{\psi }e^{2\varphi \gamma
_{5}}\psi ,  \\
\lab{eqna3}
i\gamma ^{\mu }\partial _{\mu }\widetilde{\psi } &=&m_{\psi }e^{2\varphi \gamma _{5}}\widetilde{\psi },\,\,\,m_{\psi }\equiv 4m,
\end{eqnarray}
which define the off-critical ATM model 
\br
\lab{lagrange}
\frac{1}{k}{\cal L}_{ATM}(\varphi ,\psi ,\overline{\psi })=\frac{1}{4}\partial_{\mu }\varphi \partial ^{\mu }\varphi +i\overline{\psi }\gamma ^{\mu
}\partial _{\mu }\psi -m_{\psi }\overline{\psi }e^{2i\varphi
\gamma _{5}}\psi. 
\er   

This Lagrangian is not conformal invariant and {\sl defines} the ATM model. Let us now explain the role played by the fields $\nu$ and $\eta$. In the construction of \ct{matter} these fields correspond to the extension of the $sl(2)$ loop algebra and are responsible for making the system \rf{lagrangian} conformally invariant. The field $\eta$ is a kind of conformal ``gauge connection'' and by choosing a particular constant solution ($\eta=$ const. is a solution of the equations of motion) the conformal symmetry is broken, thus obtaining a massive theory. Notice that the above gauge fixing ($\eta \rightarrow 0$ ) holds, so far, only at the clasical level. In the next section, we will give a quantum version of this reduction. On the other hand, the soliton solutions are obtained in the orbit of the vacuum solution, $\eta=0$, and their corresponding masses are determined solely by the behaviour at $x=\pm \infty$ of the space derivative of the auxiliary field $\nu$ \ct{matter}.

Let us notice that the constraints \rf{constraints}, once we have set $\eta=0$, are equivalent to 
\be
{1\o{2\pi}}\epsilon^{\mu\nu} \pa_{\nu} \, \vp=
{1\o \pi} \bar \psi \gamma^\mu  \psi.
\lab{equivcurrents}
\ee

Therefore, in the gauge fixed model, Eq. \rf{lagrange}, the Noether current \rf{noethersl2} is
proportional to the topological current \rf{topological}. This equivalence between the currents leads to some remarkable properties of the off-critical ATM model. For
instance, it implies  that the charge density $\psi^{\dagger}\psi$ is proportional to the space derivative of $\vp$; consequently, the matter field
get confined inside the $\vp$ field solitons. It can be seen, independently of the reality conditions imposed above, that the one-soliton and two-soliton solutions of \rf{lagrangian0} satisfy the relationship \rf{equivcurrents} between the Noether and topological currents \ct{bla1}.

The one-soliton solution for $\vp$ is a sine-Gordon type soliton, and the corresponding $\psi$ solution is of the massive Thirring model type \ct{orfanidis}. In addition, one can check that these solutions satisfy \rf{equivcurrents}, and so they become solution of the gauge fixed model, implying that the Dirac field must be confined inside the solitons. In fact, the one-soliton (antisoliton) type solutions of the system of equations  \rf{eqn1}-\rf{eqn5} can be written in the compact form
\br
\vp &=& 2 \arctan \( \exp \( 2m_\psi \mbox{sign}\,({e_{\psi}}) \( x-x_0-vt\)/\sqrt{1-v^2}\)\) \nonu\\
\psi &=& e^{i\theta} \sqrt{m_\psi} \, 
e^{m_\psi \mbox{sign}\,({e_{\psi}}) \( x-x_0-vt\)/\sqrt{1-v^2}}\, \(
\begin{array}{c}
\mbox{sign}\,({e_{\psi}}) \left( { 1-v\o 1+v}\right)^{1/4}  
{1 \o 1 + \mbox{sign}\,({e_{\psi}}) ie^{2 m_\psi \( x-x_0-vt\)/\sqrt{1-v^2}}}\\
-\mbox{sign}\,({e_{\psi}}) \left( { 1+v\o 1-v}\right)^{1/4}  
{1 \o 1 -\mbox{sign}\,({e_{\psi}}) ie^{2 m_\psi \( x-x_0-vt\)/\sqrt{1-v^2}}} 
\end{array}\) 
\nonu\\
\nu &=& 
 \h \log \( 1 + \exp \( 4 m_\psi \mbox{sign}\,({e_{\psi}}) \( x-x_0-vt\)/\sqrt{1-v^2}\)\) 
+ {1\o 8} m_{\psi}^2 x_{+} x_{-}
\nonu\\
\eta & = & 0 
\lab{solsimple}
\er
and the solution for $\widetilde \psi$ is proportional to the complex conjugate of
$\psi$, i.e., $\widetilde \psi\,=\, e_{\psi} \psi^{*}$ ($e_{\psi}$ being a real number). According to the reality conditions ii) or ii$^{\prime}$) imposed above, we have $e_{\psi}=\pm 1$ (for any real number $e_{\psi}$, one can make a convenient rescaling of the $\psi$ fields in order to normalize to $e_{\psi}=\pm 1$ ). In such case, the topological charge is\footnote{We are using the definition presented in \ct{bla1} for the topological charge \rf{topological} as being twice that of the reference \ct{matter}, in order to make it integer.} \ct{bla1}
\br
Q_{\rm topol.} = \mbox{sign}\,{e_{\psi}}\,=\,\pm 1.
\er

Hence, one has both the soliton and the anti-soliton solutions of the real Lagrangian sub-model, Eq. \rf{lagrangian}. Indeed, one can construct solutions of system \rf{lagrangian} starting from solutions of system \rf{lagrangian0} if the condition ii) or ii$^{\prime}$) is taken into account. In particular this will be true for the solitonic solutions. Let us mention that in \ct{bla1} the authors have discarded one of the solutions (soliton or antisoliton) since they have used only one of the possible reality conditions. Since for the study of the full ATM quantum spectrum (subsection \ref{subsec:boso}) it could be desirable to know all the soliton type classical solutions (solitons and breathers), let us mention that a carefull analysis reveals the absence of pure imaginary breather (doublet) solutions for the field $\vp$ of \rf{lagrangian0} \ct{hadron2000}; i.e., the system \rf{lagrangian} does not possess bound soliton-antisoliton pairs as a solution of its field equations. However, such type of solutions exist for a general complex but asymptotically pure imaginary $\vp$ field of \rf{lagrangian0} \ct{bla1}.

Observe that the condition \rf{equivcurrents} together with the equations
of motion for the spinor fields \rf{eqn4}-\rf{eqn5} imply the equation
of motion for $\vp$; namely, Eq. \rf{eqn1}. Therefore in the gauge fixed model,
defined by \rf{lagrange}, one can replace a second order
differential equation, i.e. \rf{eqna1}, by two first order equations,
i.e. \rf{equivcurrents}. 

The vaccum solution, $\eta=0$, was used in \ct{matter,bla1} to perform the dressing transformation in order to obtain soliton solutions, which are in the orbit of a vacuum solution. Let us emphasize that these transformations do not excite the field $\eta$ and the solitonic solutions are solutions of the gauge fixed model; i.e., the off-critical ATM model.

\subsection{Classical masses of the fundamental particles and solitons}

\label{subsec:masses}

The CATM model is conformally invariant and, therefore, their fundamental
particles are massless. Following the arguments presented in \ct{aratyn, atm}, the masses of the model are generated by the spontaneous breakdown of the
conformal symmetry of the model by the choice of a particular vaccuum
configuration $\eta =\eta _{0}=$constant. Then, considering only the linear
field approximations in \rf{poteqn1}-\rf{poteqn3} and writting the mapping $B$ as $B=e^{T}$, one gets
\br
\pa_{+}\pa_{-}T=-v_{\eta }\left[E_{-2},\left[E_{2},
T\right]\right], 
\er
\br
\pa_{+}\pa_{-}F_{1}^{\pm} =-v_{\eta }\left[E_{-2},\left[E_{2},
F_{1}^{\pm }\right]\right],
\er
where $v_{\eta }=e^{2\eta_{0}}.$ Thus, the masses of the fundamental particles are given by an eigenvalue equation
\br
\lab{eigenvalue}
\left[E_{-2},\left[E_{2},
X\right]\right]=\lambda X,
\er
where $X \in \hat{\cal{G}}_{n}$, $n=0, \pm 1$.
Since $\pa_{+}\pa_{-}=\frac{1}{4}(\pa_{t}^{2}-\pa_{x}^{2})$ we obtain the masses from the Klein-Gordon type equations, then we get
\br
\lab{particles}
m_{\varphi}=m_{\widetilde{\nu }}=m_{\eta }=0,\,\,\,\,m^{\prime}_{\psi
}=m^{\prime}_{\widetilde{\psi}}=4me^{\eta _{0}}. 
\er

Now, following the reasoning presented in Refs. \ct{matter, aratyn, atm}, we may say
that the masses of the solitons are generated by the spontaneous breakdown
of the conformal symmetry. So, the soliton mass is given by
\br
\frac{M_{\mbox{sol}}}{\sqrt{1-u^{2}}}=k\partial _{x}(\nu -\nu _{0})\mid _{-\infty
}^{+\infty }.
\er

We simply give the mass of the one-soliton (one-antisoliton) computed in \ct{matter} (for $\eta=0$); viz.,
\br
\lab{solitonmass}
M_{\mbox{sol}}=8km. 
\er

The soliton solutions are created by the eigenvalues $V$ of $E_{\pm 2}$ \ct{matter, bla1}. Expanding $V$ as $V= \sum_{n} V^{(n)}$, where $\left[Q_{s},V^{(n)}\right]=nV^{(n)}$, one observes that $\left[E_{-2},\left[E_{2},
V^{(n)}\right]\right]\, \propto\, V^{(n)}$. Then, if some $V^{(n)}, n=0,\pm 1$, does not vanish, it implies that $V^{(n)}$ must be one of the eigenvectors $X$ in \rf{eigenvalue}. In this way, we associate a soliton with a fundamental particle. In addition, the masses of the corresponding soliton and fundamental particle are determined by the same eigenvalue and they are proportional to a $U(1)$ charge \ct{matter}. 
It is then possible to have more than one fundamental particle associated to a one-soliton solution if the expansion of $V$ contains more than one non-vanishing $V^{(n)}, n=0, \pm 1$. It is just this argument that allowed us to impose the reality conditions above, identifying $\widetilde{\psi}$ as being the complex conjugate of $\psi$ (up to a factor $e_{\psi}$), since both elementary particles ( $\psi$ and $\widetilde{\psi}$) have the same mass $4me^{\eta_{o}}$, Eq. \rf{particles}, and are associated to the same soliton(antisoliton) solution \rf{solsimple} for the $\vp$ field. Therefore, the imposition of the reality condition $\widetilde{\psi}=e_{\psi}\psi^{*}$ on the CATM spinors does not spoil the particle-soliton duality, i.e., the off-critical and physically well defined ATM model inherits from the CATM the remarkable particle-soliton duality property. The above arguments indicate some sort of particle-soliton duality in the theory similar to the electromagnetic duality of some four dimensional gauge theories \ct{montonen, vw}.
 
To see more closely the role played by the sine-Gordon and
massive Thirring models in describing some aspects of the soliton (particle) sector
of our model, let us write the following suggestive relationship between the one-(anti)soliton
solution for the scalar $\varphi $ and the corresponding classical Dirac field $\psi$ solution
\begin{equation}
\lab{classicalboso}
\psi _{R}\widetilde{\protect\psi }_{L}=\frac{m_{\psi}}{4i}(e^{-2
\varphi }-1),\,\,\,\,\psi _{L}
\widetilde{\psi }_{R}=-\frac{m_{\psi}}{4i}(e^{2\varphi }-1).
\end{equation}

These relations are a good example of the classical correspondence between the sine-Gordon equation and the massive Thirring model \ct{orfanidis}. Substituting conveniently the relations \rf{classicalboso} in the equations of motion \rf{eqna1}-\rf{eqna2} (written with $\eta=\eta_{0}=$const.) one gets 
\begin{equation}
\lab{sg}
\partial ^{2}\varphi =-2m_{\psi}^{2}e^{\eta _{0}}\sin 2
\varphi
\end{equation}
and
\begin{equation}
i\gamma ^{\mu }\partial _{\mu }\psi =2m_{\psi}
\psi -4(\overline{\psi }\gamma _{\mu }
\psi )\gamma ^{\mu }\psi.
\lab{th}
\end{equation}

The equation \rf{sg} is the sine-Gordon equation and the one-(anti)soliton solution
\rf{solsimple} satisfies this equation for $\eta _{0}=\log 2,$ and \rf{th} is
the equation of motion of the massive Thirring model with coupling constant $%
g=4$ and mass M$_{Th}=m^{\prime}_{\psi}=2m_{\psi}$; of course, this $\psi $ field mass coincides with the fundamental particle mass \rf{particles}.

\section{Quantum aspects of the model}
\label{sec:quantum}

Our aim in this section is to consider the quantum aspects of the real Lagrangian sub-model defined by the Lagrangian \rf{lagrangian}. In the next subsection we consider the quantum version of the reduction CATM $\rightarrow$ ATM through the BRST analysis. In subsection \ref{subsec:boso}, we use bosonization techniques to study the model, and perform a further reduction imposing the currents (Noether and topological) equivalence as a constraint. Moreover, we discuss the quantum spectrum of the successively reduced models and outline some remarkable properties concerning the duality soliton/particle, as well as some confinement mechanism present in the ATM.

\subsection{BRST symmetry and spontaneous breakdown of conformal symmetry}
\label{subsec:brst}

We are interested in studying how the classical reduction of the model, by setting $\eta =0$,
is recovered at the quantum level. This procedure resembles the conformal affine Toda (CAT)$\rightarrow$ affine Toda (AT) reduction \ct{bonora}. We present firstly a naive version of the reduction procedure based on the path integral
\br
{\cal Z}&=&\int {\cal D}\varphi {\cal D}\overline{\psi }{\cal D}\psi {\cal D}\nu {\cal D}\eta \exp [iS(\varphi ,\psi,
\overline{\psi },\nu ,\eta )],
\er
with $S$ the corresponding action of the model defined by the lagrangian \rf{lagrangian}. In the above equation we have not written the gauge fixing term (this term should fix the left-right continuous symmetry \rf{leri1}-\rf{leri2}), the relevant ghost field and its integration measure. In \ct{bla} a term of the form $\lambda_{\mu}(\epsilon^{\mu\nu} \pa_{\nu} \, \vp-2 \bar \psi \gamma^\mu  \psi)$ ($\lambda_{\mu}$ are Lagrange multipliers), which is the equivalence between the Noether and topological currents, has been considered as a gauge fixing of the symmetry \rf{leri1}-\rf{leri2}.

We observe that $\nu$ appears as a Lagrange multiplier.
Integrating over $\nu $ and  $\eta$ successively, we get
\br
{\cal Z}=\int {\cal D}
\varphi {\cal D}\overline{\psi }{\cal D}\psi \frac{1}{\det
\partial ^{2}}\exp \left( iS(\varphi ,\psi ,\overline{
\psi })\right) , 
\er
where
\br
\lab{effective}
S(\varphi ,\psi ,\overline{\psi })=\frac{1}{k}\int
d^{2}x\Big\{\frac{1}{4}\partial _{\mu }\varphi \partial ^{
\mu }\varphi +i\overline{\psi }\gamma ^{\mu
}\partial _{\mu }\psi -m_{\psi }
\overline{\psi }e^{2i\varphi\gamma _{5}}\psi-\frac{1}{8}m_{\psi }^{2}
\Big\}.
\er

Since the determinant $\mbox{det}\, \pa^{2}$ is a constant we have derived an effective theory which is, nothing but, the affine Toda model coupled to the matter field (ATM). 

In Ref. \ct{hadron2000}, to end up with the quantum off-critical affine Toda model coupled to the matter (ATM), the ideas presented in \ct{bonora} have been used to perform a reduction process by eliminating the degrees of freedom associated to the fields $\eta$ and $\nu$. The breaking of the conformal invariance has been made by choosing a specific vacuum in the framework of perturbative Lagrangian approach. The main idea in this process was hamiltonian reduction and spontaneous symmetry breaking of conformal invariance.   

A more rigorous analysis of the reduction CATM $\rightarrow$ ATM can be made by means of BRST analysis. Following similar steps presented in the case of the $sl(2)$ affine Toda model \ct{zhang}, we add to the action \rf{lagrangian} the following ghost term
\br
S_{ghost}=i \int
d^{2}x\Big\{\h \pa_{\mu}\bar{c}\pa^{\mu}c  
\Big\} 
\er
where $c (\bar{c})$ is an anticommuting field. One can show that $S_{tot} = S + S_{ghost}$ is invariant under the BRST transformation  
\br
\lab{brst}
\d \nu =ic,\,\,\,\d \bar{c} =\eta,\,\,\,\d c =\d \eta=0,\,\,\,\d \Phi(i) =0.\,\,\,
\er
where $\Phi(i)$ denotes collectively the fields $\vp, \psi$ and $\bar{\psi}$. Then
\br
\d S_{tot}\,=\,0
\er

In addition to the equations of motion for the $\Phi(i)$ fields, we have
\br
\Box c\,=\,\Box \bar{c}\,=\,0.
\er

The conjugate momenta of the (anti)ghost are   
\br
\pi_{c}\,=\,\frac{i}{2}\pa_{t}\bar{c},\,\,\,\,\pi_{\bar{c}}\,=\,\frac{-i}{2}\pa_{t}c, 
\er
and the relevant canonical comutation relations are 
\br
\lab{canonical}
\Big\{ \Pi_{\Psi},\,\Psi \Big\}_{\mp}\,=\,-i\d(x-y), 
\er
where $\Psi$ denotes collectively the set of fields $ \{\Phi(i), \eta, \nu, c, \bar{c}\}$, and the ``$+(-)$'' signs are valid for the set of fields $\{\psi, \overline{\psi}, c, \bar{c}\}$ and $\{\Phi(i), \eta, \nu\}$, respectively.

The BRST charge is 
\br
Q_{BRST}\,=\,i \int dx (ic\,\pi_{\nu}+\eta\,\pi_{\bar{c}})\,=\,\int dx\, \left(\eta\, (\pa_{t}c)-(\pa_{t}\eta)c\right),
\er
and generates the BRST transformations \rf{brst}
\br
\lab{brst1}
\,\d \Psi =\Big[Q_{BRST},\,\Psi \Big]_{\mp}, 
\er
and satisfies the nilpotency $Q_{BRST}^{2}\,=\,0$. Moreover, considering the hermicity property of the fields $\Psi^{\dagger}=\Psi$,\,  we have $Q_{BRST}^{\dagger}=Q_{BRST}$.

Introduce a suitable wave packet system of massless particles 
\begin{eqnarray}
&\Box f_{k}(x)=0,&\\
i \int dx f^{*}_{k}(x)\vec{\pa}_{t}f_{l}(x)\,=\,\d_{kl},&\,\,&\,\,\sum_{k} f_{k}(x)f^{*}_{k}(y)\,=\,\d (x-y),
\end{eqnarray}
where $f\vec{\pa}_{t}g=f({\pa}_{t}g)-({\pa}_{t}f)g$. Therefore, since $\eta, c $ and $\bar{c}$ are free fields, we can expand them as follows  
\begin{eqnarray}
\nonu
\eta(x)\,=\,\sum_{k} \eta_{k}f_{k}(x)+\eta_{k}^{\dagger}f^{*}_{k}(x),\\
\lab{expansion}
c(x)\,=\,\sum_{k} c_{k}f_{k}(x)+c_{k}^{\dagger}f^{*}_{k}(x),\\
\nonu
\bar{c}(x)\,=\,\sum_{k} \bar{c}_{k}f_{k}(x)+{\bar{c}_{k}}^{\dagger}f^{*}_{k}(x).
\end{eqnarray}

Besides, the field $\nu$ is not a simple pole field, as can be seen from its equation of motion 
\br
\Box \nu\,=\,\h m_{\psi}^{2}e^{2\eta}-2m_{\psi}\bar{\psi}e^{\eta+2i\vp\g_{5}}\psi
\er

That is to say, the field $\nu$ is multipole field and thus can not be expanded in the simple form \rf{expansion}. Notice that, in section 4.1, precisely the asymptotic behavior of this field has been used to obtain the classical soliton masses.  Nevertheless, we can write the multipole field in the form \ct{kugo}
\br
\nu(x)\,=\,\sum_{k} \nu_{k}f_{k}(x)+\nu_{k}^{\dagger}f^{*}_{k}(x)+...
\er
where the ellipsis corresponds to the possible modes of the fields $\Phi(i)$ and $\eta$ which from the BRST transformations Eq. \rf{brst} commute with $Q_{BRST}$, and then they will not be important in our considerations. 

In terms of the creation an annihilation operators, we have
\br
Q_{BRST}\,=\,i\sum_{k} (c_{k}^{\dagger} \eta_{k}-\eta_{k}^{\dagger}c_{k}).
\er

From the canonical commutation relations \rf{canonical} we have  
\begin{eqnarray}
&&\Big[\eta_{k},\,\nu^{\dagger}_{l} \Big]\,=\,\Big[\nu_{k},\,\eta^{\dagger}_{l} \Big]\,=\,\d_{kl}\\
&&\left\{c_{k},\,\bar{c}^{\dagger}_{l} \right\}\,=-\,\left\{\bar{c}_{k},\,c^{\dagger}_{l} \right\}\,=\,-i\d_{kl}.
\end{eqnarray}

Using these relations in \rf{brst1} give us  
\begin{eqnarray}
\nonu
&&\left\{Q_{BRST},\,\bar{c}_{k} \right\}\,=\,\eta_{k},\,\,\,\Big[Q_{BRST},\,\nu_{k} \Big]\,=\,i c_{k}\\
&&\lab{ghostnumber}
\left\{Q_{BRST},\,c_{k} \right\}\,=\,0,\,\,\,\Big[Q_{BRST},\,\eta_{k} \Big]\,=\,0.
\end{eqnarray}

In addition, $Q_{BRST}$ commutes with the modes associated with the fields $\Phi(i)$.

From the above considerations we realize that the fields \{$\eta, \nu $\} and \{$c, \bar{c} $\} form a pair of BRST doublets of Kugo-Ojima's and then we may use their quartet mechanism.
The free property of the ghost fields $c$ and $\bar{c}$ implies that the total state vector space $\cal{V}$ can be decomposed persistently into a direct product ${\cal{H}}_{\Phi (i)} \otimes {\cal{H}}_{\eta, \nu} \otimes {\cal{H}}_{c, \bar{c}}$, where ${\cal{H}}_{\Phi (i)} $ correponds to the vector space spanned by the modes associated with the fields $\Phi (i)$, ${\cal{H}}_{c, \bar{c}}$ is the Fock space spanned by $c$ and $\bar{c}$ alone, and similarly for ${\cal{H}}_{\eta, \nu}$ . The proof that the ghosts decouple from the physical subspace goes as follows. Consider the following projection operator $P^{(0)}$,  
\br
P^{(0)}: {\cal H}_{\Phi(i)} \otimes {\cal H}_{\eta,\nu} \otimes {\cal H}_{c,\bar{c}} \rightarrow  {\cal H}_{\Phi(i)} \otimes |0>_{\eta,\nu} \otimes|0>_{c,\bar{c}}
\er
where the vacua $|0>_{\eta,\nu}$ and $|0>_{c,\bar{c}}$ are defined by
\br
\eta_{k}|0>_{\eta,\nu}\,=\,\nu_{k}|0>_{\eta,\nu}\,=\,0,\,\,\,\,c_{k}|0>_{c,\bar{c}}\,=\,\bar{c}_{k}|0>_{c,\bar{c}}\,=\,0.
\er

Then the projection operator commutes with $Q_{BRST}$ trivially because of the commutativity of $Q_{BRST}$ with the $\Phi(i)$ modes. Here we follow closely the procedure presented in Refs. \ct{zhang, kugo, bershadsky}. We introduce a set of operators $P^{(n)} (n\geq 1)$ defined inductively as
\br
P^{(n)}\,=\,\frac{1}{n}\sum_{k}\left(-\nu_{k}^{\dagger}P^{(n-1)}\eta_{k}-\eta_{k}^{\dagger}P^{(n-1)}\nu_{k}-ic_{k}^{\dagger}P^{(n-1)}\bar{c}_{k}+i\bar{c}_{k}^{\dagger}P^{(n-1)}c_{k}\right).
\er
These operators $P^{(n)} (n\geq 0)$ commute with $Q_{BRST}$. In addition, it can be seen that they are complete
\br
\sum_{n\ge 0}P^{(n)}\,=\,{\bf 1}, 
\er
and for $n\geq 1$, $P^{(n)}$ is BRST exact
\br
P^{(n)}&=&\left\{Q_{BRST},\,R^{(n)}\right\},\\
R^{(n)}&=&-\frac{1}{n}\sum_{k}\left(-\bar{c}_{k}^{\dagger}P^{(n-1)}\nu_{k}+\nu_{k}^{\dagger}P^{(n-1)}\bar{c}_{k}\right). 
\er

Let $|\psi>$ be a physical state in the Hilbert space $\cal{H}$. It must satisfy the physical condition
\br
Q_{BRST}|\psi>\,=\,0.
\er
Therefore any physical state $|\psi>$ annihilated by $Q_{BRST}$ is written as
\br
|\psi>\,=\,\sum_{n\ge 0}P^{(n)}|\psi>\,=\,P^{(0)}|\psi>+Q_{BRST}\left(\sum_{n\ge 1}R^{(n)}|\psi>\right)
\er
 
This means that the physical state is equivalent to its projection onto ${\cal{H}}_{\Phi (i)} \otimes |0>_{\eta,\nu} \otimes |0>_{c,\bar{c}}$\, modulo the BRST operator. That is to say, we end up with a theory in which only the modes of the fields $\Phi (i)$ are present (up to the zero modes of the \{$\nu, \eta$\} and \{$c, \bar{c}$\} fields). Therefore the physical Hilbert space of the CATM \rf{lagrangian} becomes exactly the one of the off-critical ATM model \rf{lagrange}. Then we can consider the latter as a Hamiltonian reduced and spontaneously conformal-symmetry-broken version of the CATM. 

A perturbative Lagrangian viewpoint has also been considered in order to understand the above reduction process \ct{hadron2000}. If one considers scattering processes whose external legs  consist only of $\vp, \psi$ and $\bar{\psi}$ particles, then one can easily see from the structure of the propagators and the interaction terms, the $\nu$ and $\eta$ modes decouple and therefore we can ignore the fields $\nu$ and $\eta$ in the Lagrangian \rf{lagrangian} by setting these fields to zero in the corresponding action. Then we end up with the action associated to the Lagrangian \rf{lagrange} which is just the off-critical model (ATM). However, this analysis relies upon the use of perturbation theory around the trivial vaccum, i.e., $\vp=\eta=\psi=\bar{\psi}=0,\, \nu=1/8 m_{\psi}^{2} x_{+}x_{-}$. Then the analysis of the soliton sector is missing. We believe that for this sector the zero modes of the pair of BRST doublets and the nontrivial boundary conditions of the $\nu$ and $\vp$ fields must be taken into account.

In connection to this reduction process, notice that the argument used in section 4.1 to define the masses of the solitons was to consider the field $e^{2\eta}$ as a kind of Higgs field, since it not only spontaneously breaks the conformal symmetry, but also because its vacuum expectation value sets the mass scale of the theory \ct{matter, bla1}. The role played by the non-trivial boundary conditions of the field $\nu$ in connection to the ATM model was pointed out in section 4.1. It was used to provide a relation for the classical soliton masses of the theory.

\subsection{Bosonization approach and quantum reduction}
\label{subsec:boso}

A startling property which was exploited in the study of two-dimensional field theories is related to
the possibility of transforming Fermi fields into Bose fields, and vice
versa (for a complete review of the most important references in the field
see \ct{stone}). The existence of such a transformation, called
bosonization, provided in the last years a powerful tool to obtain nonperturbative information
of two-dimensional field theories.

Even though our aim is to solve the model \rf{effective} exactly, one
may use the well known semiclassical methods to verify the validity  of the
conservation laws, such as \rf{noethersl2} and \rf{chiral}, at the quantum level; as well
as to quantize the static classical solutions. In particular, the charge
fractionization phenomenon \ct{jack} for the fermions of the model \rf{effective},
interacting with the external soliton field $\varphi,$ has been
considered in \ct{mignaco}, and the conservation laws \rf{noethersl2} and \rf{chiral} have been
verified in the semi-classical approximation.

In \ct{naon} the author derived the path-integral version of Coleman's proof
of the equivalence between the massive Thirring model and sine-Gordon models
\ct{coleman}. A Lagrangian of the type \rf{effective} (plus a free massless scalar field) appears in that
process, as a total effective Lagrangian, which gives an equivalent generating
functional, after suitable field redefinitions, of the massive Thirring model. Instead of presenting details of that calculation, we will proceed directly bosonizing the Dirac fields of \rf{effective}, which is more suited for our purposes.

Let us write the relevant action of \rf{effective} in a form which is more convenient for the bosonization of the fermion bilinears
\br
\nonu
 S_{ATM}&=&\int d^{2}x\{\frac{1}{2}\partial _{\mu
}\varphi \partial ^{\mu }\varphi +i\overline{\psi }\gamma ^{\mu }\partial
_{\mu }\psi -m_{\psi } [\overline{\psi }\frac{(1+\gamma _{5})}{2}
\psi e^{i\sqrt{8/k}\, \varphi }+\\
\lab{witten}
&&\overline{\psi }\frac{(1-\gamma _{5})}{2}\psi
e^{-i\sqrt{8/k}\,\varphi }\}, 
\er
where the field rescalings $\vp \rightarrow \sqrt{2/k}\, \vp$ and $\psi \rightarrow \sqrt{1/k}\, \psi$ have been made.  The model \rf{effective}, originally proposed by Kogut and Sinclair \ct{kogut}, has been discussed in Refs. \ct{bla1, witten} and some of the points that follow have been treated in those papers.

Introduce a new boson field representation of fermion bilinears as 
\begin{equation}
i\overline{\psi }\gamma ^{\mu }\partial _{\mu }\psi =\frac{1}{2}(\partial
_{\mu }\phi)^{2},
\end{equation}
\begin{equation}
\overline{\psi }(1\pm \gamma _{5})\psi =\mu \exp (\pm i\sqrt{4\pi }\phi),
\end{equation}

\begin{equation}
\overline{\psi }\gamma ^{\mu }\psi =-\frac{1}{\sqrt{\pi }}\epsilon ^{\mu \nu
}\partial _{\nu }\phi, 
\end{equation}
then the action now becomes 
\begin{eqnarray}
S_{B}&=&\int d^{2}x\{\frac{1}{2}\partial _{\mu
}\varphi \partial ^{\mu }\varphi +\frac{1}{2}(\partial _{\mu }\phi)^{2}
\nonu
\\
&&-\frac{1}{2}\mu m_{\psi
}[\exp i[\sqrt{8/k}\,\varphi +\sqrt{4\pi }\phi]+\exp -i[\sqrt{8/k}\,\varphi +\sqrt{4\pi }
\phi]]\}.  
\end{eqnarray}

Introducing new fields 
\begin{equation}
\widetilde{c}\equiv \frac{\varphi/a +\sqrt{4\pi }\phi}{\sqrt{4\pi +1/a^{2}}},\qquad \widetilde{
\sigma }\equiv \frac{-\phi/a+\vp \sqrt{4\pi }}{\sqrt{4\pi+1/a^{2}}},
\end{equation}
then the action takes the form 
\begin{equation}
\lab{bosonized}
S_{sG+\widetilde{\sigma}}=\int d^{2}x\Big[\frac{1}{2}(\partial _{\mu }\widetilde{c})^{2}
+\frac{1}{2}(\partial _{\mu }\widetilde{\sigma })^{2}+\mu m_{\psi }\cos ( \sqrt{4\pi+1/a^{2}}\, \widetilde{c})\Big],
\end{equation}
where $a^{2}\equiv k/8$.  

Therefore we obtain a
sine-Gordon model for the field $\widetilde{c}$ and a free, massless scalar $\widetilde{\sigma }$ field. Let us mention that an analog of the field $\widetilde{\sigma }$ will be associated in section \ref{sec:color} to the color charge sector of the full $QCD_{2}$ Lagrangian, and that of the field $\widetilde{c}$ to color singlet states (baryons and mesons) in that model.

The sine-Gordon spectrum is known exactly (see, e.g., Refs. \ct{rajaraman, dorey}). Thus, since in \rf{bosonized} $\sqrt{4\pi+1/a^{2}}$ is grater than $\sqrt{4\pi}$, it has only a
massive fermion (the soliton) and a massive antifermion (the antisoliton). There are no bound states (see below a discussion regarding this issue).

Moreover, the currents \rf{chiralcur} (setting $\eta =0$, as required in the ATM case) can be expressed as 
\br
{\cal J}=\frac{-i}{\beta \pi }\partial _{+}\widetilde{\sigma },\qquad 
\overline{{\cal J}}=\frac{-i}{\beta \pi }\partial _{-}\widetilde{\sigma }. 
\er

Let us consider a physical state $\mid \psi>$ such that, $\partial _{\pm }\widetilde{\sigma }$ $\mid \psi >=0,$ or equivalently,
\br
\lab{reductionj}
{\cal J}\mid \psi >=\overline{
{\cal J}}\mid \psi >=0.
\er

In terms of the original fields the last relations become
\br
\lab{equivalence11}
 (\epsilon ^{\mu \nu }\partial _{\nu }\varphi
-2i\overline{\psi }\gamma ^{\mu }\psi )\mid \psi >=0.
\er
 
Then, Eq. \rf{equivalence11} represents the quantum version of the equivalence between the Noether and topological currents. Therefore, their expectation values for any physical states are proportional. This is a kind of ``first quantize and then reduce'' method (see, e.g., \ct{cheche}) and becomes a quantum version of the classical reduction process, ATM $\rightarrow$ sine-Gordon model (massive Thirring) performed in \ct{bla}, imposing the equivalence of those currents as a constraint. Thus one can argue
that the relations \rf{reductionj} or equivalently \rf{equivalence11} will hold, in particular, for the energy states in the
soliton sector, i.e., the quantum solitons. It corresponds to having none of the $
\widetilde{\sigma }$ field modes excited \ct{bla1}. This reduction of the extended ``Hilbert'' space by the conditions \rf{reductionj} forbids states with certain quantum numbers. As we will explain in the next section, a $\widetilde{\sigma }$ type field is associated to the ``color'' degrees of freedom in the bosonized QCD$_{2}$ (with one flavour and two colors). 

The presence of the physical fermions (zero chirality particles) can be
clarified introducing a new fermion field that has the quantum numbers of
the physical particles. We simply introduce a new fermion field $\chi$ with 
\br
-\frac{1}{\sqrt{\pi}}\epsilon ^{\mu \nu }\partial _{\nu }\widetilde{c}= \overline{{\chi}}\gamma ^{\mu } \chi. 
\er

According to the standard rules which identify the sine-Gordon theory to the
charge-zero sector of the massive Thirring model\footnote{It has been shown \ct{klassen} that there is no full equivalence between them. They contain a subset of fields with identical correlation functions, but each model also has fields the other one does not.}, \rf{bosonized} can now be written as
\begin{equation}
\lab{chi}
S_{Th + \widetilde{\sigma}}=\int d^{2}x[i\overline{\chi}\gamma
^{\mu }\partial _{\mu }\chi-m_{F}\overline{\chi }\chi-\frac{1}{2}g(\overline{\chi}\gamma ^{\mu }\chi )^{2}+\frac{1}{2}(\partial _{\mu }\widetilde{\sigma })^{2}], 
\end{equation}
where 
\begin{equation}
\frac{\b^{2}}{4\pi}=\frac{1}{1+g/\pi },\,\,\,\,\,\, \b \equiv \sqrt{4\pi+1/a^{2}} 
\end{equation}

In this form, it is obvious that our theory consists of a massive fermion
with self-interaction, and a free massless scalar. These are the physical fermion $\chi $ and
antifermion  $\overline{\chi}$. Then the original fermions have acquired masses. For the sine-Gordon theory it is thought that the
leading order quantum correction of the soliton mass is exact and there are
no higher order corrections. It would be interesting to compute the relevant quantum corrections to the soliton mass \rf{solitonmass}, for example using the method of \ct{dashen}, of course in this case that computation should take into account the fermion fluctuations.

The breather masses in terms of the soliton mass $M_{\mbox{\small{sol}}}$ are  
\begin{equation}
\lab{breathers}
M_{N}=2M_{\mbox{\small{sol}}}\sin \left(\frac{\pi }{2}\frac{N}{\frac{8\pi}{\beta^2}-1}\right),\;\;\;N=1,2,...<\frac{8\pi}{\beta^2}-1.
\end{equation}

These quantum corrections were found by Dashen, Hasslacher and Neveu \ct{dashen} via a semiclassical quantization of the two soliton solutions, and is thought to be exact.
Let us point out that in Refs. \ct{bla1, hadron2000} the authors have not found classical (real) breather solutions for the field $\vp$. This fact reflects at the quantum level and the absence of the corresponding breather spectrum is deduced from \rf{breathers}; i.e.,  since $\b^{2}$ is greater than $4\pi$ then there are no breathers. 

In the sine-Gordon theory it is well known that the fundamental
particle is, in the quantum theory, the ground state of the breather
spectrum \ct{rajaraman}. It desappears from the spectrum when the coupling constant becomes strong enough that the elementary particle mass reaches the kink-antikink threshold. Then one can argue that the `elementary' boson associated to the field $\vp$ of our model is missing in the quantum spectrum\footnote{However, the origin of solitonic excitations due to purely quantum effects, with no soliton at the classical level, has recently been reported \ct{graham}}.

Another point which deserves a remark is the following. The theory \rf{witten} presents the chiral symmetry \rf{chiralglobal} and despite this symmetry the physical fermion particles have zero chirality.   

In fact, the original chiral current, in terms of the fields $\widetilde{c}$ and $\widetilde{\sigma}$, takes the form
\begin{equation}
A_{\mu }=\frac{i}{2}\partial _{\mu }{\widetilde\sigma }.
\end{equation}

Thus the chiral current \rf{chiral} involves only $\widetilde{\sigma} $ and not $\widetilde{c}$.  This means that the
field $\widetilde{c}$ and therefore also the physical fermion and
antifermion associated with this field $\chi$ and $\bar{\chi}$ of \rf{chi} are neutral under chirality. Then
even though the elementary fermion field $\psi$ has nonzero chirality, the
physical fermion particle has zero chirality.

\section{Realization of color symmetry in the bosonized QCD$_{2}$}
\label{sec:color}

It is reasonable to consider the ATM model a sort of $1+1$ dimensional bag model for QCD. This connection can be made evident if we introduce the two-by-two matrix $U=\mbox{exp}\{2\vp \g_{5} \}$; thus, we may rewrite the Lagrangian \rf{effective} as
\br
\nonu
\frac{1}{k}{\cal L}_{ATM}(\varphi ,\psi ,\overline{\psi })&=&\frac{1}{16}\Big\{\mbox{tr}\left[U^{-1}\pa_{\mu}U\frac{1+\g_{5}}{2}U^{-1}\pa^{\mu}U\right]-\frac{1}{2}\mbox{tr}\left[U^{-1}\pa_{\mu}U\right]\mbox{tr}\left[U^{-1}\pa^{\mu}U\right]\Big\}+\\
\lab{effectiveqcd}
&&
i\bar{\psi}\g_{\mu}\pa^{\mu}\psi-\bar{\psi}U\psi,
\er
which is a two-dimensional analog of the low energy effective Lagrangian for QCD (see, e.g. \ct{chodos}).

In the following steps we will undertake the study of the relationship between the affine Toda model coupled to the matter (ATM) and the low-energy affective Lagrangian of QCD in $1+1$ dimensions. Although QCD$_{2}$ is not exactly soluble still some interesting results may be obtained (see e.g. \ct{alves, updated} and references therein). A relation of the dynamics of QCD$_{2}$ at large distances and small energies to some massive integrable models would lead to the exact (not semi-classical) solution of the strong coupled QCD$_{2}$ as anticipated, for example, in Ref. \ct{sonnenschein}. Concerning the last point, in QCD$_{2}$ with $N_{c}$ colors and $N_{f}$ flavors it is possible to study the low-energy spectrum of hadrons in various approximation schemes. The $1/N_{c}$ expansion reveals a spectrum valid in the weak coupling regime, while the strong coupling offers the posibility of understanding the baryon as a generalized sine-Gordon soliton. In this regard, it is appealing to consider the bosonized form of the ATM model, Eq. \rf{bosonized}, as a low energy  effective Lagrangian for QCD$_{2}$. In Ref. \ct{bla1}, it was outlined a confining mechanism of some degrees of freedom in the ATM model inside the baryons of its spectrum, and it was suggested an analogous behaviour with what one expects to happen in QCD. 

With that motivation, we shall now study the bosonized QCD$_{2}$ with two internal symmetries, one flavor ($N_{f}=1$) and two colors ($N_{c}=2$). Our treatment closely follows that of Ref. \ct{ferrando2} and, therefore, we only sketch a few essential steps.
First of all, let us consider a single massless Dirac theory with two internal symmetries ($N_{f}$=1 and $N_{c}$=2) defined by
\br
\lab{massless}
{\cal L_{F}}\,=\, \frac{1}{2} \bar{\psi}_{\a}i\g^{\mu}\pa_{\mu}\psi_{\a},\,\,\,\,\,\a=1,2.
\er

This Lagrangian can be written in terms of the bosonic variables, $\vp_{\a},\,\, \a=1,2$, one for each internal fermionic degree of freedom. The bosonised lagrangian is
\br
{\cal L_{B}}\,=\, \frac{1}{2}\pa_{\mu}\vp_{\a}\pa^{\mu}\vp_{\a}, \,\,\,\a=1,2.
\er   
with the fermionic bilinear verifying
\br
:\bar{\psi}_{\a}\g^{\mu}\psi_{\a}:\,=\,-\frac{1}{\sqrt{\pi}} \epsilon^{\mu\nu}\pa_{\nu}\vp_{\a}.
\er

Considering the linear combinations $\widetilde{\sigma}=(1/2)(\vp_{1}-\vp_{2})$ and $\widetilde{c}=(1/2)(\vp_{1}+\vp_{2})$ allow us to write the $U(1)$ flavor and the third component of color isospin as topological currents,
\br
J_{U(1)}^{\mu}=\frac{1}{2}\sum_{\a}:\bar{\psi}_{\a}\g^{\mu}\psi_{\a}: =-\frac{1}{\sqrt{2\pi}} \epsilon^{\mu\nu}\pa_{\nu}\widetilde{c},\\
J_{3}^{\mu}=\sum_{\a}\mbox{Tr}:\bar{\psi}_{\a}\g^{\mu}\frac{\tau_{3}}{2}\psi_{\a}: =-\frac{1}{\sqrt{2\pi}} \epsilon^{\mu\nu}\pa_{\nu}\widetilde{\sigma},
\er
and the corresponding conserved charges as
\br
\lab{barnumber}
B=\frac{1}{\sqrt{2\pi}}[\widetilde{c}(+\infty)-\widetilde{c}(-\infty)],\\
\lab{colnumber}
T_{3}=\frac{1}{\sqrt{2\pi}}[\widetilde{\sigma}(+\infty)-\widetilde{\sigma}(-\infty)].
\er

The bosonized Lagrangian presents an explicit decoupling of flavor and color degrees of freedom when written in terms of the new fields
\br
{\cal L_{B}}\,=\, \frac{1}{2}\pa_{\mu}\widetilde{\sigma}\pa^{\mu}\widetilde{\sigma}+\frac{1}{2}\pa_{\mu}\widetilde{c}\pa^{\mu}\widetilde{c}.
\er

This Lagrangian shows explicitly the $U(1)_{F} \otimes U(1)_{C}$ symmetry. In order to realize the global $SU(2)$ symmetry, the $\widetilde{\s}$ must take non-trivial boundary conditions (NTBC), these NTBC determine the vacuum structure of the theory and the quantum numbers associated to the physical states determined from \rf{colnumber} \ct{ferrando2, halpern1}.

One can consider a mass term in \rf{massless}, which after bosonization takes the form
\br
\lab{massterm}
S_{M}&=& \int d^2x\, \hat{k}^2 \mu M (\mbox{cos}\sqrt{2\pi}\,\widetilde{c}) (\mbox{cos}\sqrt{2\pi}\,\widetilde{\s}),
\er
$\mu$ is a renormalization mass and $\hat{k}$ is a constant appearing in the bosonization formula \ct{ferrando2}. Since the classical potential has a minimum at\, $V_{M}(\widetilde{c}_{o},\widetilde{\s}_{o})=-\hat{k}^2 \mu M$, one has a degenerate vacuum formed by the $(\widetilde{c}_{o},\widetilde{\s}_{o})$ lattice 
\br
\lab{lattice}
\emptyset &\equiv&\{(n,m);\, n,m \in {\em Z} \}\, \bigcup\, \{(n+\h ,m+\h);\, n,m \in {\em Z} \}.
\er

The minima are related to the possible boundary conditions, and therefore give us precise information about the solutions of the bosonized theory, viz.,
\br
\lab{ntbc}
\lim_{x\rightarrow \pm \infty}(\widetilde{c}(x),\widetilde{\s}(x))&=&(\widetilde{c}_{o},\widetilde{\s}_{o}) \in \emptyset.
\er

Moreover, the full QCD$_{2}$ bosonized Lagrangian acquires a term due to the quark-gluon interaction \ct{baluni} 
\br
V(\widetilde{\s})&=&\frac{e^2}{32 \pi} \widetilde{\s}^2 + \sqrt{\pi} \hat{k}^4 \mu^2 [1- \frac{\mbox{sin}\sqrt{2\pi}\,\widetilde{\s}}{\sqrt{2\pi}\,\widetilde{\s}}],
\er
where $e$ is the quark-gluon interaction coupling constant. This potential is positive definite and has an unique absolute minimum $V(\widetilde{\s})=0$, at $\widetilde{\s}=0$. Therefore the interaction with the gauge field reduces the possible quantum numbers of the physical states to 
\br
\lab{lattice1}
\emptyset_{\mbox{b}} &\equiv&\{(n,0);\, n \in {\em Z} \}.
\er
 
In Ref. \ct{ferrando2}, it was verified, by computing $<T^2>$, that these states are color singlets. It is possible to give a topological interpretation to this mechanism. Once the confinement interaction is added, the vacuum structure $\emptyset$ collapses into a colorless subset of the lattice points, $\emptyset_{\mbox{b}}$; i.e., the topology associated to color becomes trivial. 

Since there is a relation between the minima structure and NTBC's, Eq. \rf{ntbc}, let us study the constraints induced by $V(\widetilde{\s})$ into the full QCD$_{2}$ action
\br
\lab{fullaction}
S&=&\int \Big[ \frac{1}{2}\pa_{\mu}\widetilde{\sigma}\pa^{\mu}\widetilde{\sigma}+\frac{1}{2}\pa_{\mu}\widetilde{c}\pa^{\mu}\widetilde{c}+\hat{k}^2 \mu M (\mbox{cos}\sqrt{2\pi}\,\widetilde{c}) (\mbox{cos}\sqrt{2\pi}\,\widetilde{\s})-V(\widetilde{\s}) \Big].
\er

Notice that the full potential has a lower bound
\br
\lab{full}
V(\widetilde{c},\widetilde{\s})&=& V_{M}(\widetilde{c},\widetilde{\s})+V(\widetilde{\s}) \geq -\hat{k}^2 \mu M, \,\,\,\forall\, (\widetilde{c},\widetilde{\s}).
\er

The equality is saturated only for the set of points defined by the lattice, Eq. \rf{lattice1}, which correspond to the color singlet states. The presence of the color interaction has transformed the {\sl free} vacuum, Eq. \rf{lattice}, in an essential way. The topologically non-trivial solutions associated to the color field $\widetilde{\s}$ are pushed to infinite energy (infinite mass) and thus the relevant Fock space has been reduced to that of only $T_{3}=0$ states (only the solutions satisfying $\widetilde{\s}(\pm \infty)=0$ can remain with finite energy \ct{ferrando2}, and  color solitons associated to non-trivial topology are given infinite energy, see Refs. \ct{ellis} for details). The singlet spectrum is constituted by particles of baryon number $B=n$, i.e., by baryons $(n=1,2,...)$ or antibaryons $(n=-1,-2,...)$ and by mesons $(n=0)$. 

In order to relate these results to the bosonized ATM model, Eq. \rf{bosonized}, we must consider the dynamics of QCD$_{2}$ at large distances and small energies. The procedure followed is very simple: we eliminate the color field $\widetilde{\s}$ from the Lagrangian via its classical equation of motion for solutions, in an infinite mass scale limit \ct{casalbuoni}, and for solutions of vanishing ``color'' charge, as discussed above; which in physical terms means that the mass scale ($M'$) must be much bigger than the color singlet state baryon (soliton) mass $M_{\mbox{\small{sol}}}$ \footnote{In solving for the $\widetilde{\s}$ field from the equations of motion, we are considering the cos$(\sqrt{2\pi}\widetilde{\s})$ expansion up to the second order term, then our approximation is meaningful in the weak field approximation. This amounts to cosidering  soliton like solutions, where the field $\widetilde{\s}$ asymptotically vanishes as $x\rightarrow \pm\infty$ and makes a lump within a range $|x|<\epsilon$, for a given $\epsilon$. }. The heavy color field $\widetilde{\s}$ decouples; as argued above, it freezes down at the value $\widetilde{\s}=\widetilde{\s}_{o}=$constant (the leading order term in the $1/M'$ expansion), leaving the physical singlet states described by the ``hadron'' field $\widetilde{c}$. Thus the system is described by the effective Lagrangian involving only the field  $\widetilde{c}$
\br
\lab{low}
{\cal L}_{\mbox{eff}}&=&\frac{1}{2}\pa_{\mu}\widetilde{c}\pa^{\mu}\widetilde{c}+\hat{k}^2 \mu M(\mbox{cos}\sqrt{2\pi}\,\widetilde{\s}_{o}) (\mbox{cos}\sqrt{2\pi}\,\widetilde{c})
\er

The spectrum of this effective Lagrangian involves baryons (a soliton and an antisoliton) and some number of soliton-antisoliton bound states (for the case we are interested in there are just two bound states or mesons). The full QCD$_{2}$ Lagrangian \rf{fullaction} has a ``hadron'' spectrum which according to recent studies would not be stable in the mesonic sector (see, e.g., Refs. \ct{alves, moha}), revealing a non-integrable character of QCD$_{2}$. We believe that, in spite of this fact (the breakdown of integrability in the mesonic sector), it is reasonable to describe the baryonic sector of \rf{fullaction} by a QCD-motivated integrable field theory, as is evident in its low energy effective Lagrangian, Eq. \rf{low}, whose color singlet baryonic sector and some of its features may be described by the ATM model, Eq. \rf{bosonized}, supplied with a condition on its state space; namely, the equivalence between the topological and Noether currents, Eq. \rf{reductionj}, imposed as a constraint on the physical states. The above idea is further supported considering that in QCD the integrable sector and the mesonic sector are seemingly decoupled \ct{decoupled}. The reduction process ATM $\rightarrow$ sine-Gordon model, which is performed imposing the equivalence between the Noether and topological currents as a constraint, is the mechanism required to describe the confining phase in the ATM model, as we shall explain in the next section.

\section{Topological Confinement in the ATM model}
\label{sec:topo}

We want to examine the question of confinement of the ``color'' degrees of freedom associated to the field $\widetilde{\s}$ in the ATM model. We will resort to a ``semi-classical'' analysis \ct{abdalla1, gross}. This goes as follows. We put a pair of external probe color charges $e^{\prime}$ and $\bar{e}^{\prime}=-e^{\prime}$ at $L/2$ and $-L/2$. In order to evaluate the inter-charge potential, we compare the ground state expectation value of the Hamiltonian associated to \rf{bosonized} in the presence of an external $\bar{e}^{\prime}e^{\prime}$ source, with the corresponding obtained in the absence of such a source \ct{abdalla}
\br
V(L)=<\Omega_{Q}|H(L)|\Omega_{Q}>-<\Omega_{0}|H(0)|\Omega_{0}>,
\er  
where $H(0)\, (H(L))$ and $|\Omega_{0}> \,(|\Omega_{Q}>)$ are the Hamiltonian and ground state in the absence (presence) of the probe charges, respectively. Let us consider an additional term in \rf{witten} such that an external spinor field couples to $\vp$ in the same form, viz.,
\br
m_{\psi }^{\prime} [\overline{\psi}_{\mbox{ext}}\frac{(1+\gamma _{5})}{2}
\psi_{\mbox{ext}} e^{i\sqrt{8/k}\, \varphi }+\overline{\psi}_{\mbox{ext}}\frac{(1-\gamma _{5})}{2}\psi_{\mbox{ext}}
e^{-i\sqrt{8/k}\,\varphi }],
\er
the coupling constant, taken as $m_{\psi }^{\prime}$, will be associated to the mass of the external fermion. Next, one can bosonize the external fermion bilinears in the usual form,
\br
\left(\bar{\psi}\g^{\mu}{\psi}\right)_{\mbox{ext}} = -\frac{1}{\sqrt{\pi}}\epsilon^{\mu\nu} \pa_{\nu}Q
\er
where
\br
\lab{charge11}
Q=-\frac{ke^{\prime}}{2}\sqrt{\pi}\Xi,
\er
where $e^{\prime}$ is the external color charge\footnote{Notice that the factor $k/2$ is introduced in relation \rf{charge11} in order to match the definition of the charges presented in section \ref{sec:real} in terms of the unrescaled  ATM fields of the Lagrangian \rf{lagrange} or \rf{effective}.}. The field  $\Xi$ above is taken as
\br
\Xi=[\Theta(x-L/2)-\Theta(x+L/2)],
\er
this function will represent the external color charge potential of the probe charges $e^{\prime}\bar{e}^{\prime}$. This will give a vanishing total ``color'' charge. 

Introducing the field $\widetilde{c}_{Q}$ as     
\br
\widetilde{c}_{Q}=\frac{\sqrt{4\pi}Q+\vp/a}{\sqrt{4\pi+1/a^{2}}}
\er
we can write the total bosonized action as
\br
\lab{totalboso}
S_{\mbox{tot}}\,=\,\int d^{2}x[\frac{1}{2}(\partial _{\mu }\widetilde{c})^{2}
+\frac{1}{2}(\partial _{\mu }\widetilde{\sigma })^{2}+\lambda\cos (\b \, \widetilde{c})+\lambda^{\prime} \cos (\b \, \widetilde{c}_{Q})],
\er
where $\lambda \equiv  \mu m_{\psi}$ and $\lambda^{\prime}\equiv  \mu m_{\psi}^{\prime}$. 

The bosonized Lagrangian \rf{totalboso} will be the starting point for our future analysis and from this point on our treatment will be purely classical. It should, however, be noted that this is better than treating the original fermionic model, since bosonization capture some of the quantum nature of the model. Therefore, one may argue that semiclassically, the constraints \rf{reductionj} are equivalent to $\widetilde{\s}=\widetilde{\s}_{o}=$constant, and consequently
 the degrees of freedom eliminated by them correspond to the  $\widetilde{\s}$ field. 

Since we are looking for a static potential between heavy ``quarks'' in a color singlet state, the field $\widetilde{\s}$ may be set equal to a constant (zero) in accordance with the reduction process considered in section 5, Eq. \rf{reductionj} and the discussion above. This procedure is the analog of the QCD confinement mechanism, in which the topology associated to an analogous field is dynamically collapsed, as discussed in the last section. Then the static Hamiltonian, in terms of the field $\widetilde{c}$ and incorporating the external color charge potential, becomes 
\br
\nonu
H(L)&=&\int dx[\frac{1}{2}(\widetilde{c}^{\prime})^{2}-\lambda\cos ( \b\, \widetilde{c})-\lambda^{\prime}\cos (\sqrt{4\pi}Q+\frac{\widetilde{c}}{a^2\b})].
\er

The classical potential has a minimum at 
\br
V_{min}=-\lambda-\lambda^{\prime},
\er
and this implies a degenerate minima formed by the set of points 
\br
\lab{points}
\frac{2\pi m}{\b}\,\,\,\, \mbox{and} \,\,\,\, 2\pi n a^2 \b; \,\,\,\,\,\, m,n \in {\em Z}.
\er  

Then the parameters must satisfy the relationship, $a^2 \b^2 n=m$, which implies 
\br
(\frac{\pi k}{2}+1)n =m.
\er

Therefore $\pi k$ should be rational. The coupling of the external field $Q$, related to the probe charges, has not affected in an essencial way the nature of the coupling constant $k$ (a quantization of $k$ is of course expected, since the CATM is related to the one of WZNW \ct{matter}; see section $3$ and relation \rf{couplings}). The minima provide us important information about the solutions of the bosonized theory, thus allowed us, for example, to determine the correct quantum numbers ($B$ or $T_{3}$) in QCD$_{2}$. In the same way one can consider the boundary conditions in the ATM case
\br
\lim_{x\rightarrow \pm\infty} \widetilde{c}(x)\, =\,\widetilde{c}(\pm \infty),\,\,\,\,\, \lim_{x\rightarrow \pm\infty} \widetilde{\sigma}(x)\, =\,\widetilde{\sigma}(\pm \infty)(=0,\,\,\mbox{for ``color'' singlet states})
\er

The inter-charge potential energy $V(L)$ is associated to the field configurations that minimize the Hamiltonian. Then the equation of motion of the field $\widetilde{c}$ which follows from the bosonized Lagrangian \rf{totalboso} is, in the static case 
\br
\lab{nonlinear}
\widetilde{c}^{\prime\prime}-\frac{\lambda}{\b}\sin ( \b\, \widetilde{c})-\frac{\lambda^{\prime}}{a^{2}\b}\sin (\sqrt{4\pi}Q+\frac{\widetilde{c}}{a^2\b})=0.
\er

It is well known that the solution of the above classical equation of motion in the bosonic version of the theory contains quantum mechanical information from the fermionic theory. Then we need to solve \rf{nonlinear} imposing some boundary conditions. The  nonlinear equation \rf{nonlinear} is not exactly solvable and can only be solved considering the first order term in the sine expansions\footnote{This approximation is only meaningful if the conditions $\b |\widetilde{c}(\pm\infty)|<<1$ and $4 \pi e^{\prime}<<k$ are satisfied.}. In this approximation the equation reduces to
\br
\widetilde{c}^{\prime\prime}-w^2 \widetilde{c}\,=\,f\, \Xi,
\er
where 
\br
w^2\, =\,\lambda+\frac{\lambda^{\prime}}{a^4\b^2 },\,\,\,\,f=-\frac{4\pi\lambda^{\prime}e^{\prime}}{a^2\b k}.
\er

This equation is exactly solvable and its solution is
\br
\widetilde{c}\,=\,\left\{ \begin{array}{ll}Fe^{-\a x}+ \widetilde{c}(+\infty)& x > L/2,\\
Ce^{\a x}+De^{-\a x}+E &|x|<L/2,\\
Ae^{\a x}+ \widetilde{c}(-\infty)& x < -L/2.
\end{array}\right.
\er

We will consider the non-trivial boundary conditions (NTBC): $\widetilde{c}(+\infty)=\widetilde{c}_{+}\neq\, 0$ and $\widetilde{c}(-\infty) = 0$. Moreover, the solutions and their derivatives should be matched at the boundaries $\pm L/2$ giving us a set of algebraic equations for the unknown coefficients. For example the parameter $\a$ is given by
\br
\a^{2} = \lambda + \frac{\lambda^{\prime}}{a^4 \b^2}.
\er

The potential derivative becomes
\br
\frac{d}{dL}V(L)&=&\frac{d}{dL}[H(L)-H(0)]\nonu
\\
&=&\frac{2\lambda e^{\prime}}{k}\pi\Big[\sin(\sqrt{4\pi}Q(L/2)+\frac{\widetilde{c}(L/2)}{a^2\b})
-\sin(\sqrt{4\pi}Q(-L/2)+\frac{\widetilde{c}(-L/2)}{a^2\b})\Big].\nonu
\er

Expanding the sine-terms and keeping the first order terms we have 
\br
\frac{d}{dL}V(L)=\frac{2\lambda^{\prime} e^{\prime}}{k a^2 \b}\pi[\widetilde{c}(L/2)- \widetilde{c}(-L/2)].
\er

Finally, for the above field configurations which minimize the Hamiltonian we compute the inter-charge potential 
\br
\lab{todapot}
V(L)=\frac{\widetilde{c}_{+}\,\lambda^{\prime}e^{\prime} \pi}{a^2 k \b}\Big[ L + \frac{1}{\a} e^{-\a L}\Big].
\er

Thus, we observe that the inter-charge potential has both the confining and screening type terms. However, if the non-trivial boundary condition (NTBC) for the sine-Gordon soliton (baryon), $\widetilde{c}_{+}\neq 0$, is considered and the inter-charge separation increases, i.e., $L \rightarrow +\infty$, then the confining term dominates. Therefore, taking into account the particle spectrum of our model (see section 5), we can argue that the ATM model presents a permanent confining phase of the ``color'' degrees of freedom inside the baryons of the model (solitons and antisolitons).

\section{Discussions}
\label{sec:discussions}

To summarise the results of this paper: it has been shown that the baryonic sector of the low energy effective Lagrangian of QCD$_{2}$ with one flavor and two colors, and some of its properties may be described by the integrable and off-critical affine Toda model coupled to the matter field (ATM), provided the equivalence between the Noether and topological currents of the ATM model were imposed as a constraint.  

It could be worthwhile to mention that a potential of type \rf{todapot} is present in the Toda lattice system \ct{mtoda}. In the Toda lattice model a particle interacts only with its nearest neighborhoods via a potential defined by \rf{todapot}, where $L$ represents the distance between the interacting particles. The appearance of this type of potential and the connection of the ATM with the Toda lattice systems, if there exists any in the context studied above, deserves a further investigation.

We believe this model could be useful as a toy-model to implement the so called ``Cheshire Cat Mechanism'' as a gauge symmetry, used to study confinement in $1+1$ dimensional chiral bag models \ct{vento}. Work in this direction is in progress \ct{cheshire}.

It is natural to inquire into the extension of the above analysis to models constructed using affine Lie algebras associated to higher rank finite Lie algebras, since the same property, i.e., the equivalence between the Noether and topological currents appears in a class of special models defined in \ct{matter}. It could be interesting to check the validity of this property for the various soliton type solutions, and whether the currents equivalence holds true at the quantum level. As the general CATM, for any affine Lie algebra, is related to the two-loop WZNW model, the analysis could also be performed using the WZNW fields, in terms of which the matter fields can be written locally but in a somewhat cumbersome way.

\vspace{1cm}

Note Added: After the completion of this work we became aware of the Refs. \ct{zhao} where the authors were able to recover similar models to the CATM considered above. In the case of the $sl(2)$ CATM studied by us, it is the special $sl(2)$ case of a family of their so-called bosonic superconformal affine Toda models based on arbitrary affine Lie algebra. The authors considered the construction and the integrability properties, as well as the conformal correspondences of such models. It is also worth to point out that one of the main points of Ref. \ct{matter} was the uncovering of the special class of models in which the equivalence between the Noether and topological currents holds true, at least, at the classical level. 
\vspace{1cm}

\noindent {\bf Acknowledgements}

The author is grateful to Professors L.A. Ferreira, G.M. Sotkov, and  A.H. Zimerman for valuable discussions. I thank Professor L. Zhao for correspondence and letting me know the Refs. \ct{zhao}. R. Bent\'{\i}n and C. Tello are also akcnowledged for fruitful conversations. I would like to thank FAPESP for financial support. 

\vspace{1cm}

\appendix
\section{Notations and Conventions}
\label{appa}

We use the following conventions in two dimensions. The metric tensor is $g_{\mu\nu}=\mbox{diag}(1, -1)$ and the antisymmetric tensor $\epsilon_{\mu\nu}$ is defined so that $\epsilon_{01}=-\epsilon^{01}=-1$. $\pa_{\pm}$ are derivatives w.r.t. to the light cone variables $x_{\pm}=t\pm x$. The gamma matrices are in the following representation:
\begin{equation}
\gamma _{0}=-i\left( 
\begin{array}{cc}
0 & -1 \\ 
1 & 0
\end{array}
\right) ,\qquad \gamma _{1}=-i\left( 
\begin{array}{cc}
0 & 1 \\ 
1 & 0
\end{array}
\right) , \qquad \gamma _{5}=\gamma _{0}\gamma _{1} =\left( 
\begin{array}{cc}
1 & 0 \\ 
0 & -1
\end{array}
\right),
\end{equation}
satisfying anticommutation relations 
\br
\{\gamma _{\mu },\gamma _{\nu }\}=2g_{\mu \nu }{\bf 1,} 
\er
so the spinors $\psi$ and $\bar{\psi}$ are of the form
\br
\psi =\left( 
\begin{array}{c}
\psi _{R} \\ 
\psi _{L}
\end{array}
\right) ,\qquad \widetilde{\psi }=\left( 
\begin{array}{c}
\widetilde{\psi }_{R} \\ 
\widetilde{\psi }_{L}
\end{array}
\right) , \qquad \bar{\psi }=\left( 
\begin{array}{cc}
\widetilde{\psi }_{R}\,\, 
\widetilde{\psi }_{L}
\end{array}
\right) \gamma_{0}.
\er

\section{The $sl(2)^{(1)}$ affine Lie algebra notations}
\label{appb}

Let us work with the Chevalley basis
generators $H^{n},E_{\pm }^{n}$, $D$ and $C$ of  $sl(2)^{(1)}$. The commutation relations are 
\begin{eqnarray}
\left[ H^{m},H^{n}\right] &=&2mC\delta _{m+n,0},  \nonumber \\
\left[ H^{m},E_{\pm }^{n}\right] &=&\pm 2E_{\pm }^{m+n},  \nonumber \\
\left[ E_{+}^{m},E_{-}^{n}\right] &=&H^{m+n}+mC\delta _{m+n,0},  \nonumber \\
\left[ D,T^{n}\right] &=&nT^{n},\qquad T^{n}\equiv H^{m},E_{\pm }^{n};  
\label{1}
\end{eqnarray}
all other commutation relations are trivial. The grading operator for the
principal gradation defined by the vector ${\bf s}=(1,1)$ is \ct{kac} 
\br
Q_{{\bf s}}\equiv H^{0}+N_{{\bf s}}D-\frac{1}{2N_{{\bf s}}}Tr(H_{{\bf s}
}^{2})C\,\,\,\, (N_{{\bf s}}\equiv{\bf s}_{1}+{\bf s}_{2}=2). 
\er

Then, the corresponding eigensubspaces are 
\br
\widehat{{\cal G}}_{0}=\{H^{0},C,Q_{{\bf s}}\}; 
\er
\br
\qquad \qquad \widehat{{\cal G}}_{2n+1}=\{E_{+}^{n},E_{-}^{n+1}\}\qquad n\in 
{\em Z}; 
\er
\br
\widehat{{\cal G}}_{2n}=\{H^{n}\},\qquad n\in \{{\em Z}-0\}.\qquad 
\er

Under the above gradation the affine Lie albebra decomposes as
\br
\widehat{{\cal G}}=\bigoplus_{s} \widehat{{\cal G}}_{s}
\er
with
\br
\Big[ \widehat{{\cal G}}_{s},\,\widehat{{\cal G}}_{s}\Big] \subset \widehat{{\cal G}}_{s+r}.
\er

The grades $s$ take zero, positive and negative values, i.e.,
\br
\widehat{{\cal G}}=\widehat{{\cal G}}_{+}\bigoplus \widehat{{\cal G}}_{0}\bigoplus \widehat{{\cal G}}_{-}
\er
with
\br
\widehat{{\cal G}}_{+}\equiv \bigoplus_{s>0}\widehat{{\cal G}}_{+},\,\,\,\,\widehat{{\cal G}}_{-}\equiv \bigoplus_{s<0}\widehat{{\cal G}}_{s}
\er

In addition, we will use a special basis for the generators of $\widehat{%
{\cal G}}_{0}$ such that they are all orthogonal to\ $Q_{{\bf s}}$ and $C.$
Thus, shifting the Cartan elements as $\widetilde{H}^{0}$ = $H^{0}-\frac{1}{2%
}C,$ one gets 
\begin{eqnarray}
Tr(C^{2}) &=&Tr(C\widetilde{H}^{0})=Tr(Q_{{\bf s}}^{2})=Tr(Q_{{\bf s}}
\widetilde{H}^{0})=0,  \nonumber \\
Tr(Q_{{\bf s}}C) &=&2,Tr(\widetilde{H}^{0}
\widetilde{H}^{0})=Tr(H^{0}H^{0})=2.
\end{eqnarray}

Here we have used $Tr(CD)=1,$ and the normalization for the root, $\alpha
^{2}=2$. For more details of such a special basis, see appendix $C$ of Ref.
\ct{atm}.

\end{document}